\documentclass[aps,prb,twocolumn,showkeys]{revtex4-1}
\usepackage{graphics,graphicx,epsfig,color,hyperref,latexsym,float,amsmath}
\usepackage{times}
\input pdfcolor.tex

\begin{document}

\title{Superfluid-Mott transition and thermal phases of correlated lattice bosons:\\
a classical fluctuation theory}
\author{Abhishek Joshi and Pinaki Majumdar}
\affiliation{Harish-Chandra Research Institute, HBNI, 
Chhatnag Road, Jhusi, Allahabad 211019}
\date{\today}

\begin{abstract}
We present a method that generalises the standard mean field theory of 
correlated lattice bosons to include amplitude and phase fluctuations
of the $U(1)$ field that induces onsite particle number mixing. This arises 
formally from an auxiliary field decomposition of the kinetic term in
a Bose Hubbard model.  We solve the resulting problem, initially, by 
using a classical approximation for the particle number mixing field and
a Monte Carlo treatment of the resulting bosonic model. In two dimensions
we obtain $T_c$ scales that dramatically improve on mean field 
theory and are within about 20\% of full quantum Monte Carlo estimates
at density $n=1$. 
The `classical approximation' ground state, however, is still mean field, 
with an overestimate of the critical interaction, $U_c$, for the superfluid 
to Mott transition. By further including low order quantum fluctuations 
in the free energy functional we improve significantly on the $U_c$, and 
the overall thermal phase diagram. The classical approximation based method
has a computational cost linear in system size. The methods readily 
generalise to multispecies bosons, disorder, and the presence of traps, 
and yield real frequency response functions.
\end{abstract}

\maketitle

\section{Introduction}

The experimental realization 
\cite{Bloch,Fort,Ernst,Klaus} 
of a quantum phase transition
from a superfluid to a Mott insulator for bosons in an 
optical lattice  
bridged the gap 
\cite{Immanuel Bloch,Zwerger}  between condensed matter
and cold atom physics.  
The quantitative features of the transition are captured 
by quantum Monte Carlo (QMC) simulations in both two dimensions 
\cite{Sansone} and three dimensions 
\cite{Capogrosso}.  These have  established  the phase diagram 
involving, (i)~the interaction, $U$, driven superfluid to Mott 
insulator transition in the ground state at integer filling, 
and (ii)~the thermally driven superfluid to normal Bose liquid 
transition.  

While QMC provides high accuracy numerical
results for the thermodynamic features,
one would
want methods that (i)~shed light on the key physical 
effects, and (ii)~allow access to dynamical properties.
Mean field theory \cite{Sheshadri,Fisher}, although
quantitatively not very accurate, 
highlights the crucial effect of particle number
fluctuation in the superfluid ground state.
More sophisticated methods, {\it e.g}, 
strong coupling expansion \cite{Monien,Freericks}, 
variational calculations \cite{dos}, the projection
operator technique \cite{Anirban},cluster gutzwiller\cite{Dirk}  
and variational  cluster
schemes \cite{Knap,Michael} improve on the
mean field ground state and yield results that 
agree well with
QMC for the zero temperature transition.

There is less insight into the finite 
temperature situation.
Mean field theory has a finite temperature
generalisation
but leads to a large overestimate
of  $T_c$ scales. Slave particle methods
\cite{Stoof,Lu} and a self consistent standard 
basis operator\cite{SBO} 
approach have been used to study the thermal physics.
Recently a bosonic version of dynamical mean
field theory \cite{Volhardt,Tong,Anders} 
(BDMFT) has been
developed - retaining all 
local quantum fluctuations but 
ignoring spatial correlations.
Among the methods above 
only the results of 
BDMFT compare reasonably with QMC in 
terms of thermal properties.

We present an alternate extension
of mean field theory in this paper,
with emphasis on {\it spatial
fluctuations}, which we believe are important in lower
dimensions, and include temporal (quantum)
fluctuations only approximately. 
The two versions of this approach are, borrowing from
the nuclear physics literature,
\cite{Negele,Bertsch,Attias}
(i)~the static path
approximation (SPA), and (ii)~the perturbed 
static path approximation (PSPA), for the 
functional integral defining the partition function.
We present the analytic basis of these methods and
obtain the following results for the two dimensional (2D)
Bose Hubbard model within a Monte Carlo implementation
of these schemes.

\begin{figure}[b]
\centerline{
\includegraphics[width=6.0cm,height=5.2cm,angle=0]{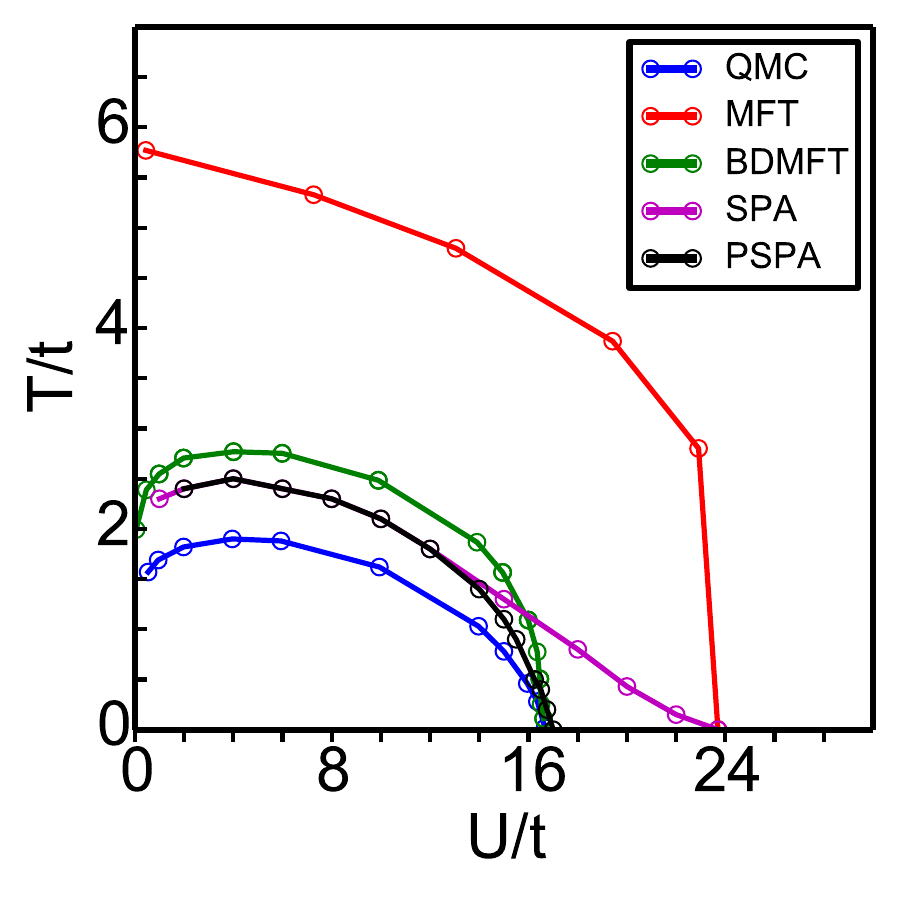}
}
\caption{Color online:
Superfluid $T_c$ from different methods for the 2D
Bose Hubbard model at filling $n=1$.  We use $U_c(0)$,
the $T=0$ critical interaction, and $T_c^{max}$,
the maximum superfluid $T_c$, to characterise each scheme.
(i)~Mean field theory (MFT) yields $U_c(0) = 24t$ and $T_c^{max} = 6t$,
(ii)~for BDMFT $U_c(0) \sim 17t$ and $T_c^{max} \sim 3t$,
(iii)~for our simplest method,
the SPA, $U_c(0) = 24t$ and $T_c^{max} \sim 2.5t$,
(iv)~for the perturbed SPA, $U_c(0) \sim 17t$ and $T_c^{max} \sim 2.5t$.
For full quantum Monte Carlo (v)~$U_c(0) \sim 17t$ and $T_c^{max} \sim 2t$.
The abbreviations are explained in the text.
}
\end{figure}

(i)~Ground state: at zero temperature 
the SPA expectedly reproduces the mean field phase
boundary between the superfluid and the Mott insulator.
PSPA results are almost indistinguishable from
full QMC. 

(ii)~Thermal scales:
at filling $n=1$ both SPA
and PSPA lead to a maximum superfluid $T_c$ that is
$T_c^{max} \sim 2.5t$ (where $t$ is the hopping scale).
Within QMC $T_c^{max} \sim 2t$  while
mean field theory predicts $T_c^{max} = 6t$, see Fig.1. 

(iii)~Number fluctuations:
we establish the distribution of the 
key `hybridisation' field that is responsible for
on site particle number fluctuations and illustrate 
spontaneous fluctuations even in the Mott phase 
at finite temperature 

(iv)~Spatial correlations:  
we extract a characteristic lengthscale $\xi(T,U)$ for
the hybridisation field. $\xi$ diverges as the thermal
transition is approached, emphasising that the hydridisation 
field follows a spatially correlated distribution - not 
accessible within `local' theories.

(v)~Amplitude-vs-phase fluctuations:
while both amplitude and phase fluctuations are relevant at small
$U$, the large $U$ low temperature problem is dominated by 
phase fluctuations - which allows us to construct and benchmark a 
XY model. This well approximates $T_c(U)$ and provides 
physical insight.

The rest of the paper is organized as follows: in Section II
we discuss our approach and its numerical implementation.
Section III discusses our results for the ground state and 
thermal behaviour. 
Section IV provides an analysis in terms of an XY model, and
discusses computational checks.

\section{Model and method}

We explore our methods in the context of the 
2D Bose Hubbard model at unity filling:
$$
H = -t\sum_{<ij>}a^{\dagger}_i a_j 
- \mu \sum_i n_i
+ {U \over 2} \sum_{i} n_i(n_i-1) 
$$
where $a$ and $a^{\dagger}$ are the usual second quantised
bosonic operators, $t$ is the nearest neighbour hopping amplitude,
$U$ is the onsite repulsion, and the 
boson density is fixed at $n=1$ by using a 
chemical potential $\mu$.
 
 In order to arrive at our approximations 
we follow a standard  path integral approach 
 \cite{Kampf,HTC,Sengupta}. 
Within the path integral formalism the 
full partition function is given by
\begin{eqnarray}
Z &=& 
\int {\cal D}b {\cal D} {\bar b}~ e^{-(S_0+S_K)}\nonumber \cr 
S_0 &=& 
\int_0^{\beta} d\tau 
[\sum_i
{\overline{b}_i}(\partial_{\tau} - \mu)b_i 
+ {U \over 2} \sum_i \overline{b}_i b_i(\overline{b}_i b_i -1)] \cr 
S_K &=& \int_0^{\beta} d\tau (-t)\sum\limits_{<ij>}
(\overline{b}_i b_j + h.c.) 
\nonumber
\end{eqnarray}
The $b$'s in the path integral are space and (imaginary) time dependent
classical fields.
$S_0$ involves the local terms and $S_K$ 
the kinetic energy.

We separate the kinetic term 
as follows:
$ S_K  =  S_K^a + S_K^b $,
\begin{eqnarray}
S_K^a & = &
-\sum\limits_{\vec{k}}
A_{\vec{k}}{\overline{b}_{\vec{k},0}} b_{\vec{k},0} \cr
S_K^b & = & -\sum\limits_{\vec{k}}
B_{\vec{k}}{\overline{b}_{\vec{k},0}}b_{\vec{k},0}-
\sum\limits_{n\neq 0,\vec{k}}t_{\vec{k}}
{\overline{b}_{\vec{k},n}}b_{\vec{k},n} 
\nonumber 
\end{eqnarray}
where 
$ A_{\vec{k}}  = \theta( t_{\vec{k}}) t_{\vec{k}} $ and
$B_{\vec{k}}  = \theta(-t_{\vec{k}})  t_{\vec{k}}$
and $t_{\vec k} = 2t(cosk_xa + cosk_ya)$.
Note that $S_K^a$ involves only zero frequency modes of 
$b_{\vec k}$, and only ${\vec k}$ for which the
tight binding energy, $-t_{\vec k}$,
is negative. $S_K^b$ involves the
rest of the contributions.

We use a Hubbard-Stratonovich transformation to 
decouple only $S_K^a$, keeping $S_K^b$ untouched.
$$
e^{-S_K^a}=\int
{\cal D} \psi {\cal D} \psi^*~e^{-\sum\limits_{\vec{k}}
\psi^*_{\vec{k}}\psi_{\vec{k}}+\sum\limits_{k}\sqrt{A_{\vec{k}}
 }(\psi_{\vec{k}}\overline{b}_{\vec{k},0}+h.c.)}
$$
The focus on $S_K^a$ to start with is to retain bounded weight
in a sampling process, as we will explain soon.
All the terms above 
can be collected  and Fourier transformed
back to real space and imaginary time to give us  
\begin{eqnarray}
Z~ &=& \int 
{\cal D} \psi {\cal D} \psi^* {\cal D} b {\cal D} {\bar b}~
e^{-(S+S^b_K)} \cr
S~ & =& S_0[b, {\bar b}] 
-\sum_{ij}(C_{ij}
{\overline{b}}_{{i}}\psi_{{j}}+hc) +\sum_{i}{\psi}_{i}^*\psi_{i}\cr
C_{ij} & =& 
\frac{1}{N}
\sum\limits_{\vec{k}}\sqrt{A_{\vec{k}}}
e^{i\vec{k}(\vec{r_i}-\vec{r_j})} 
\nonumber
\end{eqnarray}
This is an exact representation of Bose Hubbard model.
The fourfold symmetric 
real function $C_{ij}$ is shown in Fig.2, with respect to 
the reference site $i$ taken at the origin.

We examine two approximations:
(i)~Drop $S_K^b$ completely and solve for the physics arising from
$S$ - this is the SPA for the partition function and 
treats the hybridising field as classical. (ii)~Treat
the effect of dynamical fluctuations contained in 
$S_K^b$ to quadratic order, this is the PSPA.

\subsection{The static path approximation (SPA)}

We examine the SPA in detail and relegate the algebra for the
PSPA to an Appendix and only provide the final PSPA 
result here.  Within the SPA
\begin{eqnarray}
Z & = &  \int 
{\cal D} \psi {\cal D} \psi^* {\cal D} b {\cal D} {\bar b}~e^{-S} \cr
&= & 
\int {\cal D} \psi {\cal D} \psi^*  Tr[exp(-\beta H')] \cr
&= &  
\int {\cal D} \psi {\cal D} \psi^*  e^{-\beta F\{\psi\}} 
\nonumber
\end{eqnarray}
In the expression above
\begin{eqnarray}
F &  =&  \sum\limits_i F_i=-\frac{1}{\beta}
\sum_i log(Tr[exp(-\beta H_i')]) \cr
H'_i & = & 
({a}_i^\dagger\Phi_i+h.c)+ {U \over 2} n_i(n_i-1)
-\mu~n_i+{\psi}_{i}^*\psi_{i}  \cr
\Phi_i & = &\sum\limits_{j}C_{ij}\psi_j
\nonumber
\end{eqnarray}
In what follows we will call $\psi_i$ the {\it auxiliary field},
which is what we actually update, and $\Phi_i$ the 
{\it hybridisation field}.

\begin{figure}[t]
\centerline{
\includegraphics[angle=0,width=5.0cm,height=4.3cm]{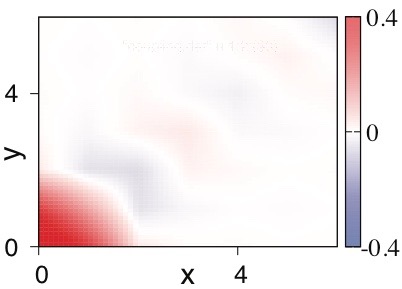}
}
\caption{Color online: The coupling $C_{ij}$ 
between the boson field $b_i$
and the auxiliary field $\psi_j$.
The reference site $i$ is taken to be
the origin $(0,0)$. The plot highlights the rapid decay of $C_{ij}$
with separation $R_{ij}$, justifying a `cluster treatment' (see
text) of the energy cost.
}
\end{figure}

The task at any temperature is 
to sample over configurations of $\psi$ with weight 
$
P[\psi]\propto\prod_i Tr [e^{-\beta H'_i}]
$.
The sampling is performed 
using the Metropolis algorithm.  
This involves calculating
$Tr[ e^{-\beta H'_i}]$, for which 
we construct the matrix for $H'_i$ 
in the local occupation number basis, 
$\sim (a_i^{\dagger})^n \vert 0 \rangle$,
truncated 
at boson occupancy $N_b = 10$.  

We use a local update scheme.
The hybridization at any
site $R_i$ is given by $ \sum_{j}C_{ij}\psi_j$,
so changing $\psi_i$ affects the $\Phi$ of
all sites on the lattice. However, 
since $C_{ij}$ falls
off quickly with distance we use $C_{ij}$
generated for a cluster of size $ N_c = 6  \times  6$,
centered around $R_i$.  The  free energy
required for the update is the sum of $F_i$ on the cluster.

The computational cost of an update is $\propto N_c N_b^3$,
where $N_c$ is the cluster size, 
and $N_b^3$ is the matrix diagonalisation cost for each site
in the cluster.  For the system sweep the cost would be
$N N_c N_b^3$, where $N$ is the system size. The Monte Carlo 
method, therefore, is ${\cal O}(N)$ with a large
prefactor.

After equilibration we store $\psi$ configurations 
to calculate thermodynamic averages.
We adjust $\mu$ to remain
at unity filling at each temperature.
At $T=0$ where every site sees the same
hybridisation SPA reduces to mean field theory. 
At finite temperature both the amplitude and
the phase of the $\psi_i$ (and so the $\Phi_i$)
fluctuate.

\subsection{Perturbed static path approximation (PSPA)}

To improve the SPA one needs to build back the
neglected quantum fluctuations.
To derive the corrected form of free energy,
we solve the $S$ part exactly 
and include 
corrections due to $S_K^b$ perturbatively
by summing diagrams with `self avoiding' paths 
to all order and then replacing the series sum
by an exponentiated term.
The full partition function
is approximated by
\begin{eqnarray}
\begin{split}
&Z\approx \int {\cal D} \psi {\cal D} \psi^* ~e^{-\beta( F+X-Y)}\nonumber\\
&F= -\frac{1}{\beta}log(Tr[exp(-\beta H')])\\
\end{split}
\nonumber
\end{eqnarray}

Here $F$ is the contribution due to $S$ part of 
the action. The details are given in the Appendix.
In the updated scheme sampling over $\psi$ configuration
is to be done with weight $P[\psi]\propto e^{-\beta(F+X-Y)}$.
$ X$  depends upon coupling $B_{ij}$. 
Since $B_{ij}$ also falls off
very quickly with distance, we use $B_{ij}$ generated for 
$4\times 4$ lattice. $X$ can be written as sum  of contribution 
coming from onsite, nearest neighbour bonds, and next nearest
neighbour bonds  for every site whereas $Y$ is sum of
contribution from only nearest neighbour bonds.

The change in the 
free energy, relevant for the update, is
computed on  a  $6 \times 6$ cluster centered on
the update site.
While calculating $Y$ 
we sum over only the lowest four eigenstates. The 
rest follows as in SPA.

While we emphasize the PSPA results in this paper,
due to its quantitative accuracy, we often compare
and contrast it to SPA - given the 
conceptual simplicity of the SPA scheme.

\begin{figure}[b]
\centerline{
\includegraphics[angle=0,width=6.7cm,height=4.9cm]{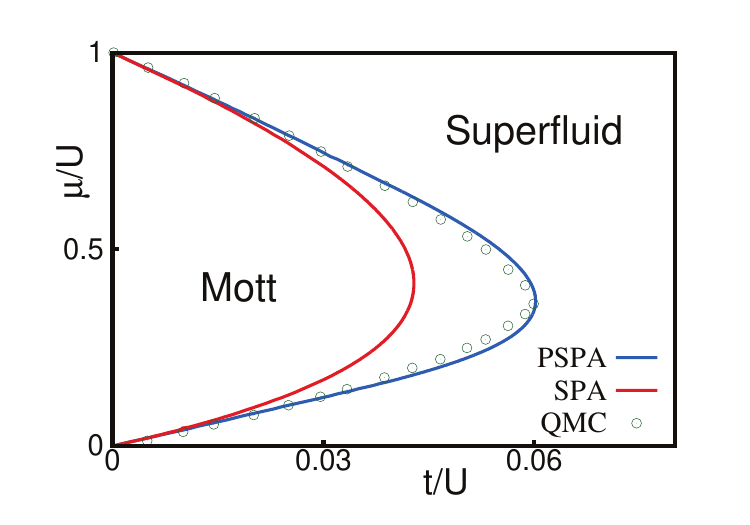}
}
\caption{Color online: The ground state of the 2D Bose Hubbard model
for varying $\mu/t$ and $U/t$. The region within the lobe
is a Mott insulator with $n=1$. The inner lobe, with a larger
$(t/U)_c$ is the result of the SPA and is the same as in mean field
theory. The outer lobe, with smaller $(t/U)_c$, is the result
of the PSPA and is indistinguishable from QMC results.  }
\end{figure}

\section{Results}

\subsection{The ground state}

Fig.3 shows the comparison between ground state obtained 
using the SPA and PSPA schemes.
The value of the uniform hybridisation 
$\Phi$ is zero inside the Mott lobe and 
increases with increase in $t/U$. The critical value
at the tip of the Mott lobe,
is found to be $(t/U)_{c} = 0.0428$ within SPA, the
same as the mean field result.
The critical point value is $(t/U)_{c} = 0.0595$ 
under PSPA, very close to the QMC value.

Fig.4 compares the SPA and PSPA energy 
functions for uniform hybridisation.
Numerical data are 
shown by open circles, at $U/t=10,~15,~16$ and $18$,
and even order polynomial fits to them are 
shown by firm lines.
The minimum of the respective function decides 
the $T=0$ order parameter within that schemes.
In the following discussion we ignore the angle 
dependence of $\Phi$ since it is irrelevant at $T=0$.

In the $U = 10t-18t$ window shown the SPA ground
state is superfluid. There is a local maximum in 
$E_{SPA}(\Phi)$ at $\Phi=0$ and a $\Phi \neq 0$ minimum
that moves to smaller value with increasing $U/t$.
The SPA
leads to a second order transition from superfluid to Mott
insulator at $U/t =24$.

For the PSPA there is always a minimum at $\Phi=0$. The
$\Phi \neq 0$ minimum is deeper for $U \lesssim 17t$
- leading to a superfluid state. 
At $U=18t$, panel (d) in Fig.4, the PSPA minimum at
finite $\Phi$ is no longer visible. There is a weakly 
first order SF to Mott transition 
within the PSPA at $U \sim 17.5t$.

\begin{figure}[b]
\centerline{
\includegraphics[angle=0,width=4.3cm,height=4.3cm]{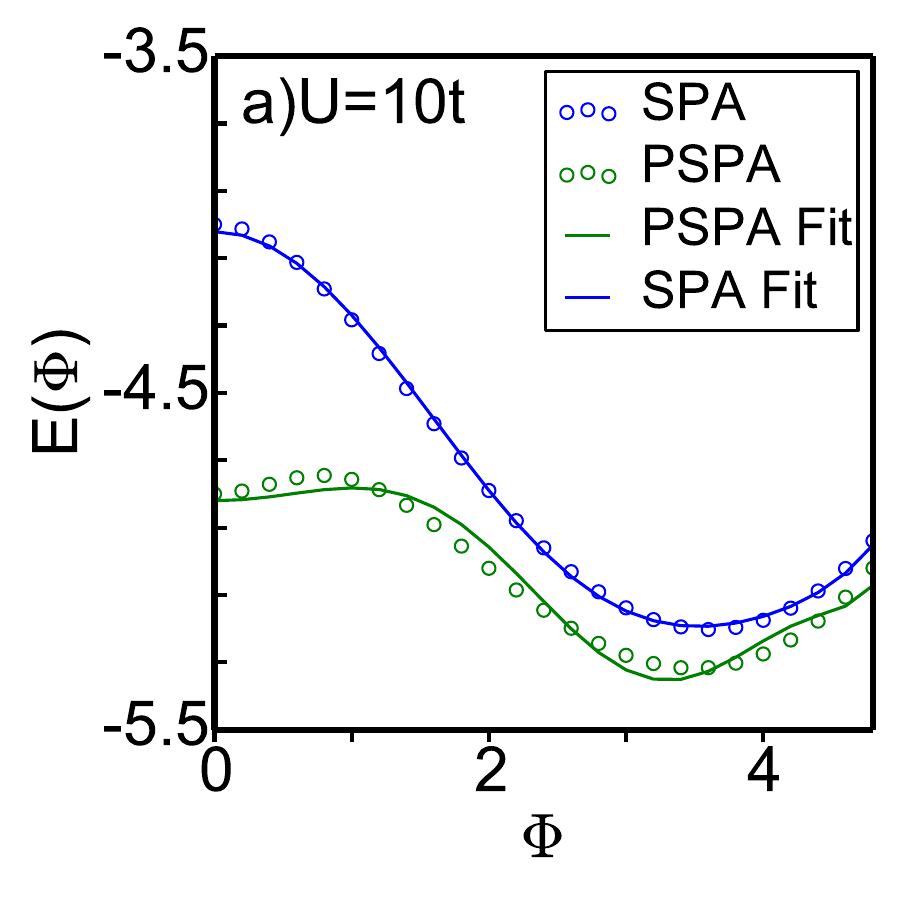}
\includegraphics[angle=0,width=4.3cm,height=4.3cm]{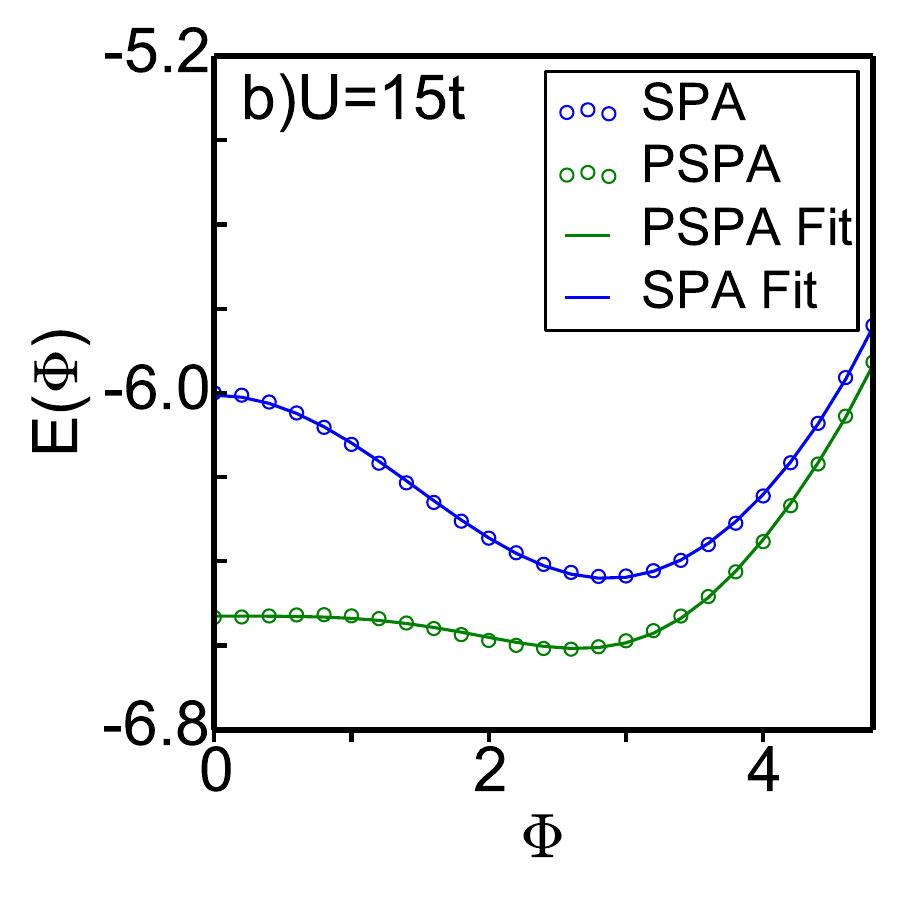}
}
\centerline{
\includegraphics[angle=0,width=4.3cm,height=4.3cm]{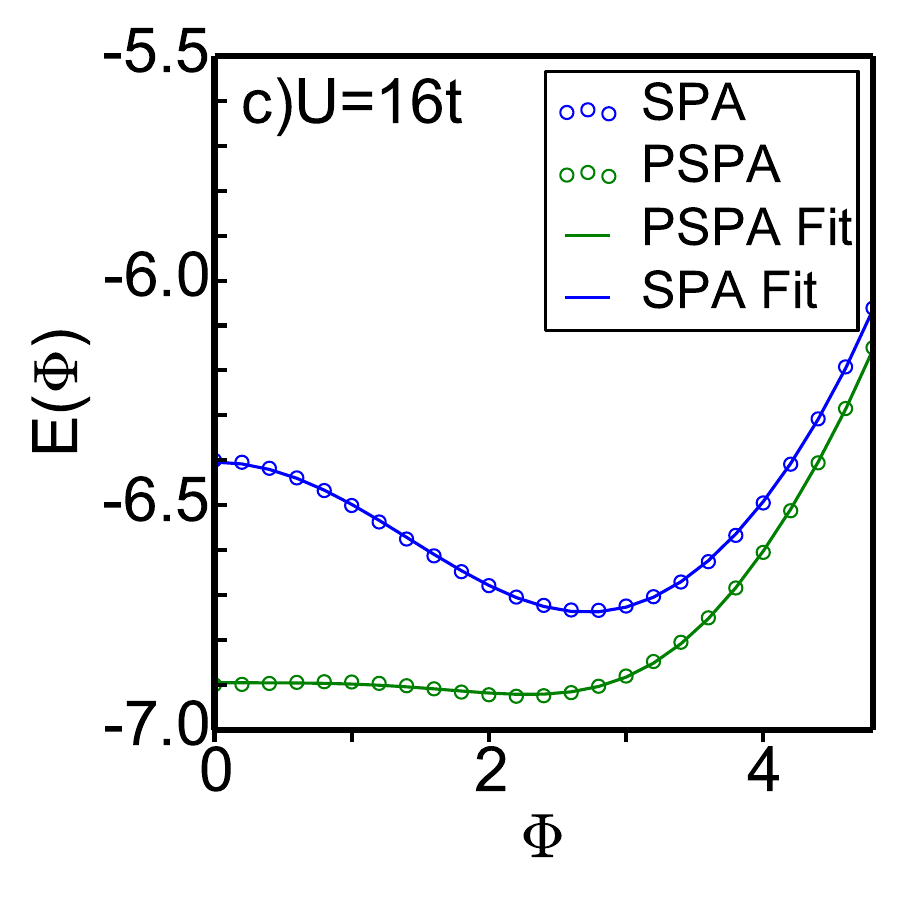}
\includegraphics[angle=0,width=4.3cm,height=4.3cm]{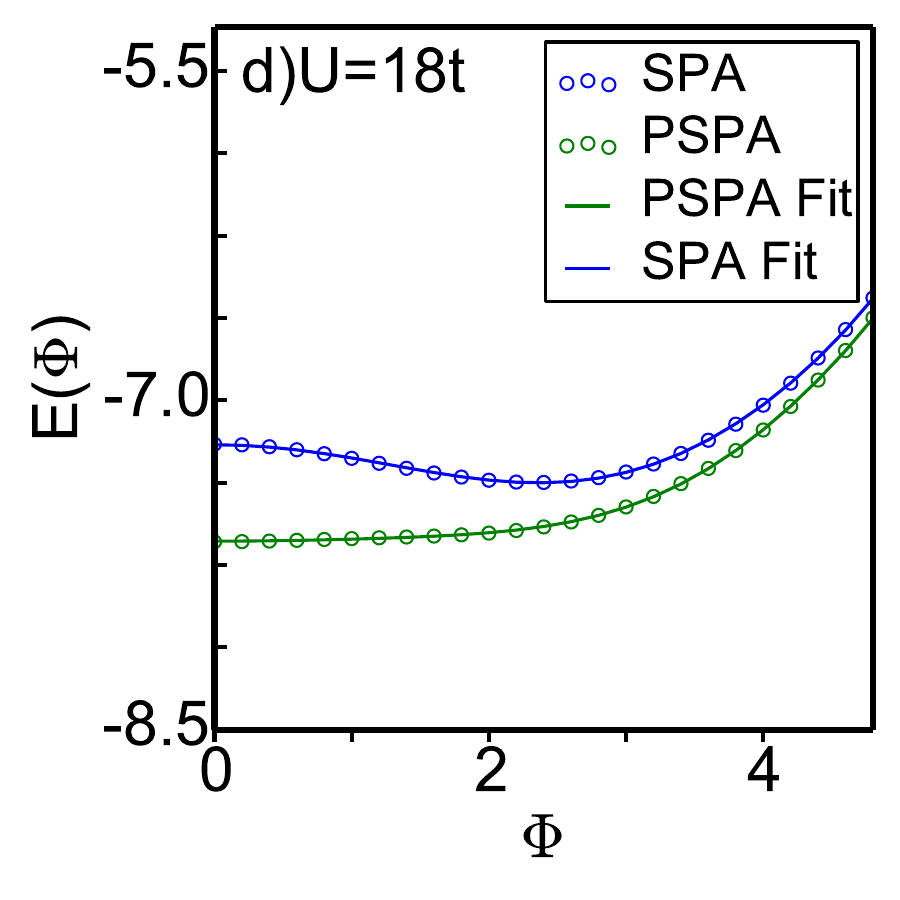}
}
\caption{Color online:
The open circles in panels (a)-(d) show the SPA and PSPA energy
functions, plotted for an uniform order parameter, $\Phi$, for the
values of $U/t$ indicated. The SPA and PSPA minima match with each
other in panels (a) and (b), in panel (c) the difference is 
noticeable, while in (d) the SPA minimum is at finite $\Phi$ while 
the PSPA minimum has shifted to zero. The firm lines are even order
Landau fits to the SPA and
PSPA functions upto $\vert \Phi \vert^6$   }
\end{figure}
\begin{figure}[t]
\centerline{
\includegraphics[angle=0,width=3.1cm,height=4.2cm]{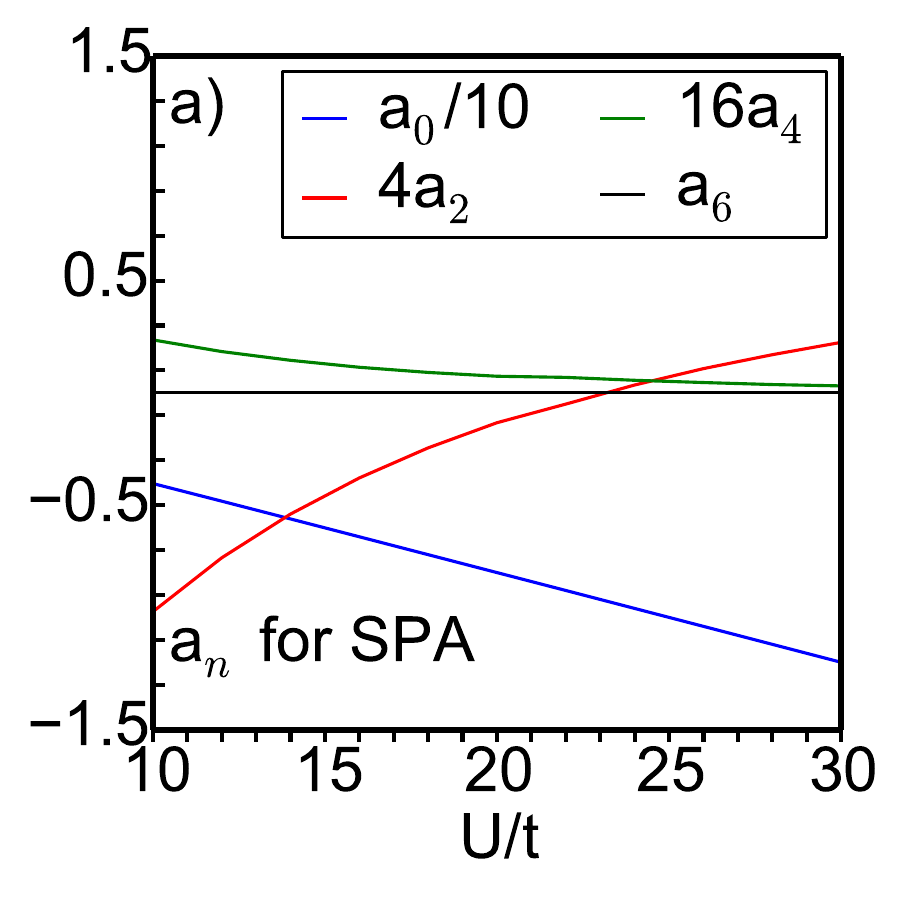}
\hspace{-.3cm}
\includegraphics[angle=0,width=3.1cm,height=4.2cm]{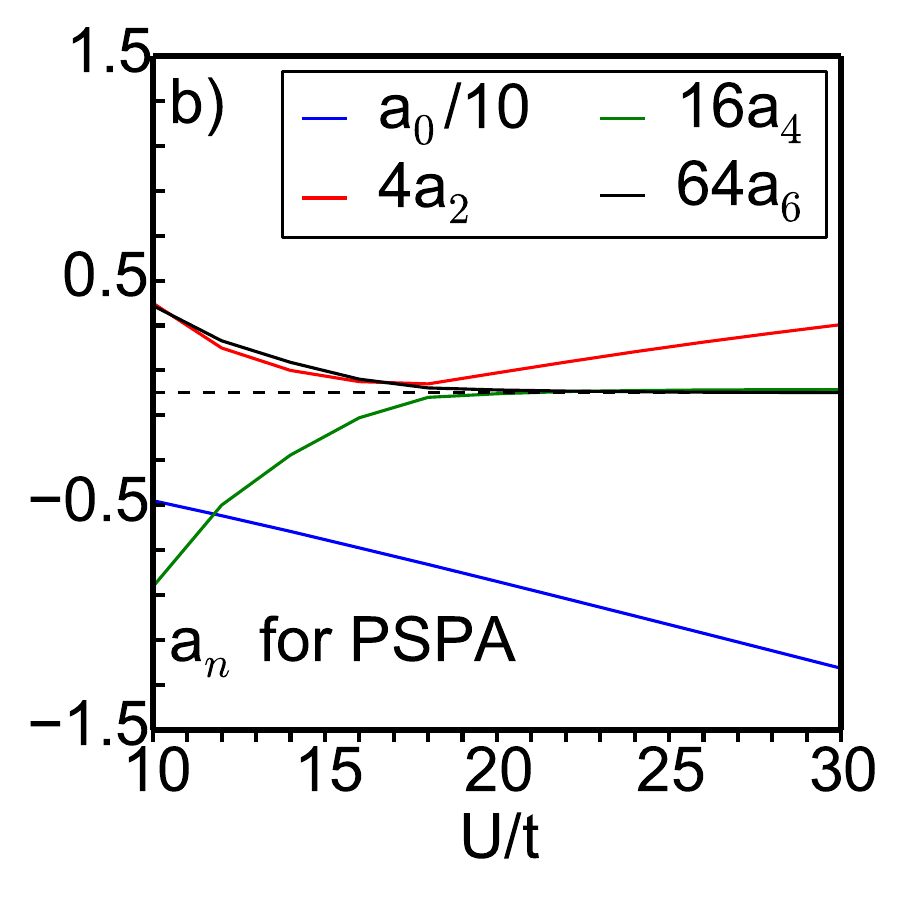}
\hspace{-.3cm}
\includegraphics[angle=0,width=2.7cm,height=4.1cm]{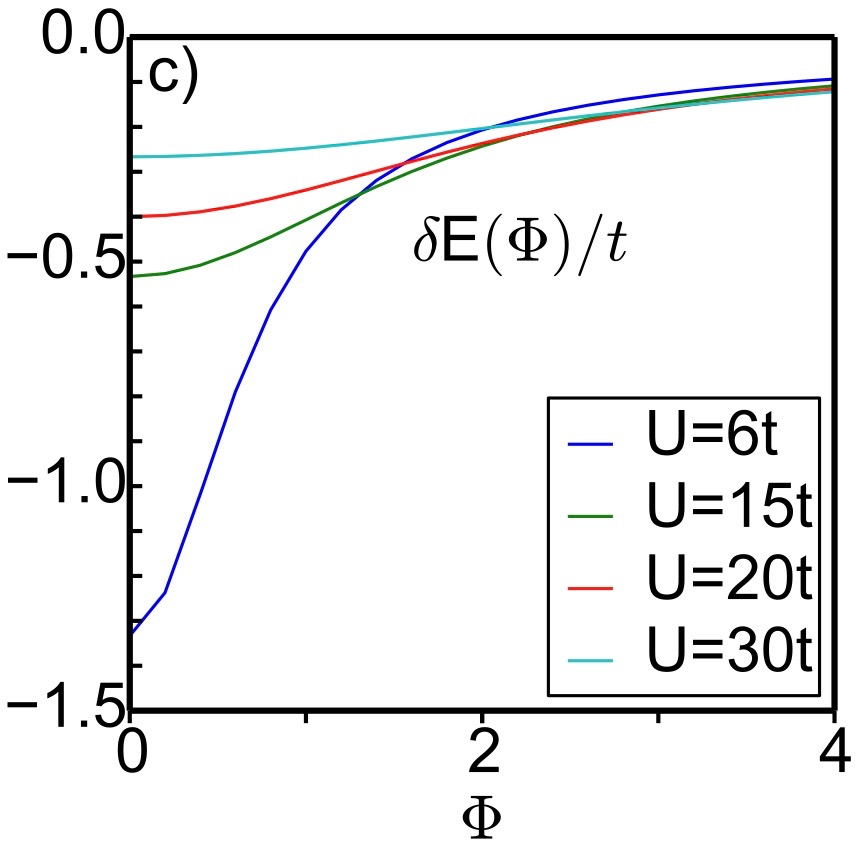}
}
\caption{Color online:
(a) Landau parameters for SPA and (b) Landau parameters for
PSPA. Note that $a_4$ remains positive for all $U/t$ in SPA,
$a_6$ is negligible, and the second order transition is driven 
by sign change in $a_2$.
For PSPA $a_2$ remains positive at all $U/t$, so there is always
a minimum at $\Phi  =0$, and the transition is driven
by $a_4$ becoming less negative with growing $U/t$. $a_6$ remains
positive. (c)~Shows the behaviour of $E_{PSPA} - E_{SPA}$, highlighting
the source of the large positive $a_2$ and the negative $a_4$ in PSPA.
}
\end{figure}

To locate the origin of the difference between SPA and
PSPA we analysed
the energy functions in terms of their Landau expansion.
The SPA energy at $T=0$ has a Landau expansion of the form: 
$
E_{SPA}(\Phi) = \sum_m a_m  \Phi^m
$
where $\Phi$ is the uniform hybridisation field. Naturally only
even powers arise in the expansion. 
The coefficients $a_m$ can be estimated as
\begin{eqnarray}
a_0 &=& {U \over 2}n(n-1)-\mu n \cr
a_2 & =& {1\over 4}+~t~G_{ii}(0) \cr
a_4 & =& -(t^2/4) G_{ii}^{2c}
\nonumber
\end{eqnarray}
where
$ G_{ii}(0)= - \int_0^{\beta}d\tau<T_\tau b(\tau)b^\dagger(0)> $
and
$$
G_{ii}^{2c}  =  \int_0^{\beta} d\tau_1 .. d\tau_3 
<T_{\tau}b(\tau_1)b(\tau_2)b^\dagger(0)b^{\dagger}(\tau_3)>-
2\beta G^2_{ii}(0)
$$
The $G_{ii}$ are correlators in the $\Phi=0$ problem.

For PSPA the expansion is similar, of the form
$
E_{PSPA}(\Phi) = \sum_m a'_m  \Phi^m
$
but an analytic derivation of the coefficients is
more involved and is described in the
Appendix.

Fig.5 shows the Landau parameters obtained by fitting 
the SPA and PSPA energy functions, 
and the energy difference function 
$\delta  E = E_{PSPA} - E_{SPA} $. 
For SPA the coefficient $a_4$ is always positive, 
$a_6$ is small and positive, 
and the SF to Mott transition is driven 
by $a_2$ changing from negative to positive. 
For PSPA $a_2 >0$ and $a_6 > 0$  throughout. 
There is always a minimum at
$\Phi $ =0. 
$a_4$ is large and negative at small $U/t$,
generating the SF minimum. With increasing $U/t$ the coefficient
$a_4 \rightarrow 0$ and the finite $\Phi$ minimum becomes higher in
energy than the $\Phi =0$ minimum. This drives a weak first order SF to
Mott transition.
Panel $(c)$ shows $\delta  E$, and the data reveals the origin of the 
difference in the $a_2$ and $a_4$ coefficients between SPA and PSPA.

We found a slight disagreement between fit parameters and 
the analytic estimate for coefficients of SPA functional: 
$a_0$ and $a_2$ match well but $a_4$ deviates when $(U_c - U) 
\gg t$.

\begin{figure*}[t]
\centerline{
\includegraphics[angle=0,width=12.0cm,height=6.0cm]{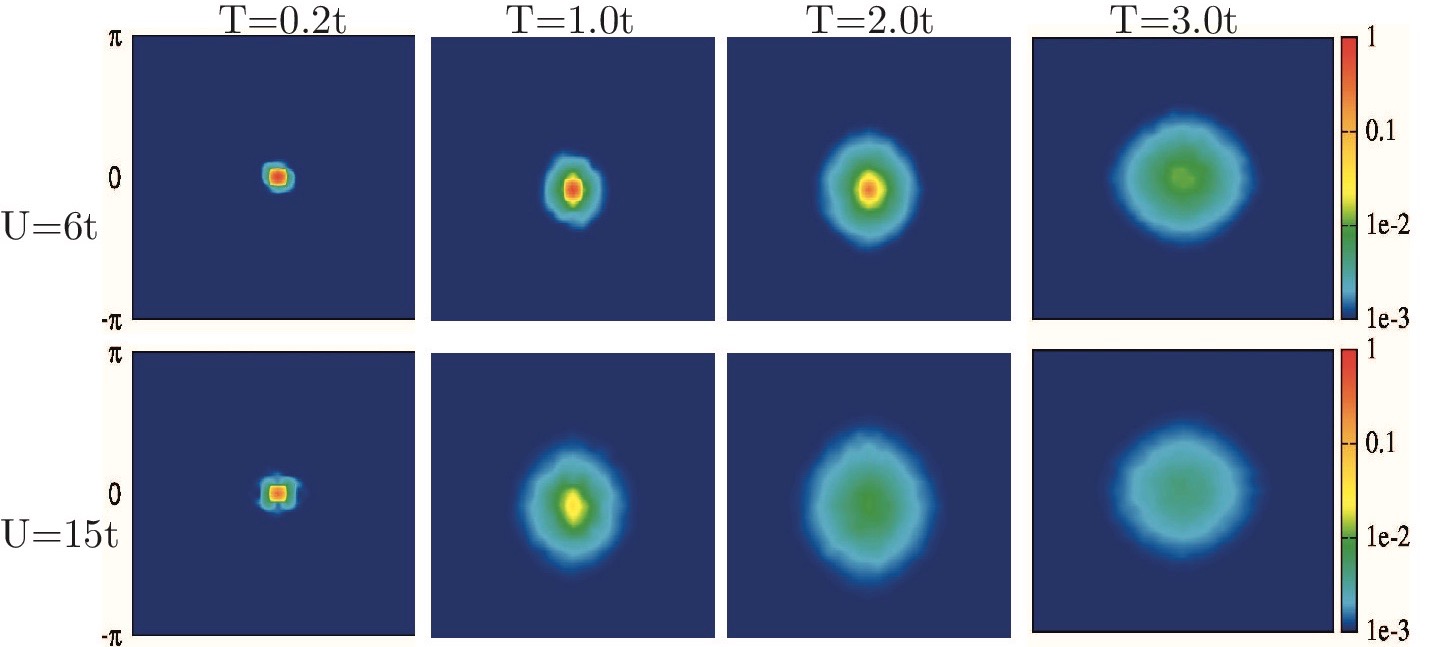}
}
\caption{Color online:
The momentum distribution, $n_{\bf k} =
\langle \langle a^{\dagger}_{\bf k} a_{\bf k} \rangle \rangle$,
for varying interaction strength and temperature in the superfluid
phase.  The intensity scale is logarithmic to simultaneously capture
the condensate peak in the superfluid state and show the particle
distribution in high temperature normal  phase.  }
\end{figure*}

\subsection{Thermal behaviour}

\subsubsection{Indicators}

We track the following indicators to locate the thermal 
transition and also quantify the amplitude and phase
fluctuations in the problem.\\
(i) The structure factor:
$
S({\bf q}) = {1 \over N^2} \sum_{ij} \Phi_i \Phi_j^* e^{i {\bf q}. 
({\vec r}_i - {\vec r}_j)} 
$
The $T_c$ is extracted from the $T$ 
dependence of the $S(0,0)$.\\  
(ii)~The momentum distribution of bosons,
\begin{eqnarray}
 \begin{split}
\langle  n(\vec{k}) \rangle
= \frac{1}{N}<\sum\limits_{i,j}e^{-i\vec{k}.(\vec{r_i}-\vec{r_j})}
Tr[e^{-\beta H'}a^\dagger_ia_j]>\nonumber
 \end{split}
\nonumber
\end{eqnarray}
(iii)~The distribution, and moments, of the hybridisation field:
$$
P(\vert \Phi \vert) = {1 \over {N N_{\alpha}}}
\sum_{\alpha, i} \delta(\vert \Phi \vert
- \vert \Phi_i^{\alpha} \vert)
$$
where $\alpha$ is a configuration label and $N_{\alpha}$ the
number of configurations averaged over. Note that $P$ is
normalised. The distribution allows
us to calculate the moments
$
\langle y^n \rangle = \int dy P(y) y^n
$
where $y = \vert \Phi \vert$. $n=1$ yields the mean,
$\vert \Phi_{av}$.  We compute the standard deviation, or `width'
of the $\vert \Phi \vert $ as $\Phi_{wid} = 
\sqrt {\langle \vert \Phi^2 \vert \rangle -
\langle \vert \Phi \vert^2 \rangle} $.\\
(iv)~The spatial correlation in a single Monte Carlo snapshot:
$$
C_i= \sum_{\delta} \vert \Phi_i \vert \vert \Phi_{i + \delta}  \vert
cos(\theta_i - \theta_{i + \delta})
$$
where $\delta$ are the nearest neighbours of $i$.\\
(v)~Finally, we compute a correlation length $\xi(U,T)$ 
from a fit to the structure factor data:
$
\xi(T,U) = \frac{1}{2sin(\pi/L)}\sqrt{\frac{S(0,0)}
{S(2\pi/L,0)}-1}
$

\subsubsection{Thermal phase diagram}

We now turn to the thermal phase diagram shown
in Fig.1 and discuss it in more detail.
The figure compares the $T_c(U)$  obtained from
SPA and PSPA with mean field theory, bosonic 
DMFT, and full QMC.
The SPA $T_c$ is already a significant 
improvement over 
mean field theory and compares reasonably with QMC 
in the intermediate $U/t$ regime. 
The $T_c$ of the superfluid should vanish as $U \rightarrow 0$
since there is no phase stiffness in the absence of interactions.
We do obtain a non monotonic dependence
of $T_c$ on $U/t$ with a maximum located at 
$ U/t \sim 4$, consistent with QMC, but the
$U/t \rightarrow 0$ limit is not captured
correctly.
The PSPA scheme leads to $T_c$'s that are close to
the SPA results for $U/t \lesssim 14$, beyond which
the PSPA $T_c$ quickly drops to zero. 
However even PSPA does not correctly capture the
asymptotic behaviour as $U/t \rightarrow 0$.
%
%

We will
show detailed results for four representative 
$U/t$ values:
(i)~$U=6t$ where the system is a moderate coupling 
superfluid, (ii)~$U=15t$ where it is a strongly interacting 
superfluid, (iii)~$U=20t$ - a `weak Mott state', just
beyond $U_c$, and (iv)~$U=30t$ - a deep Mott state.
We  highlight spatial maps and distributions at four
temperatures, $ T=0.2t, t, 2t$ and $3t$.   
We first discuss the two superfluid regimes in the next
subsection, and the two Mott states in the subsection after.

\begin{figure}[b]
{
\includegraphics[angle=0,width=2.7cm,height=4.0cm]{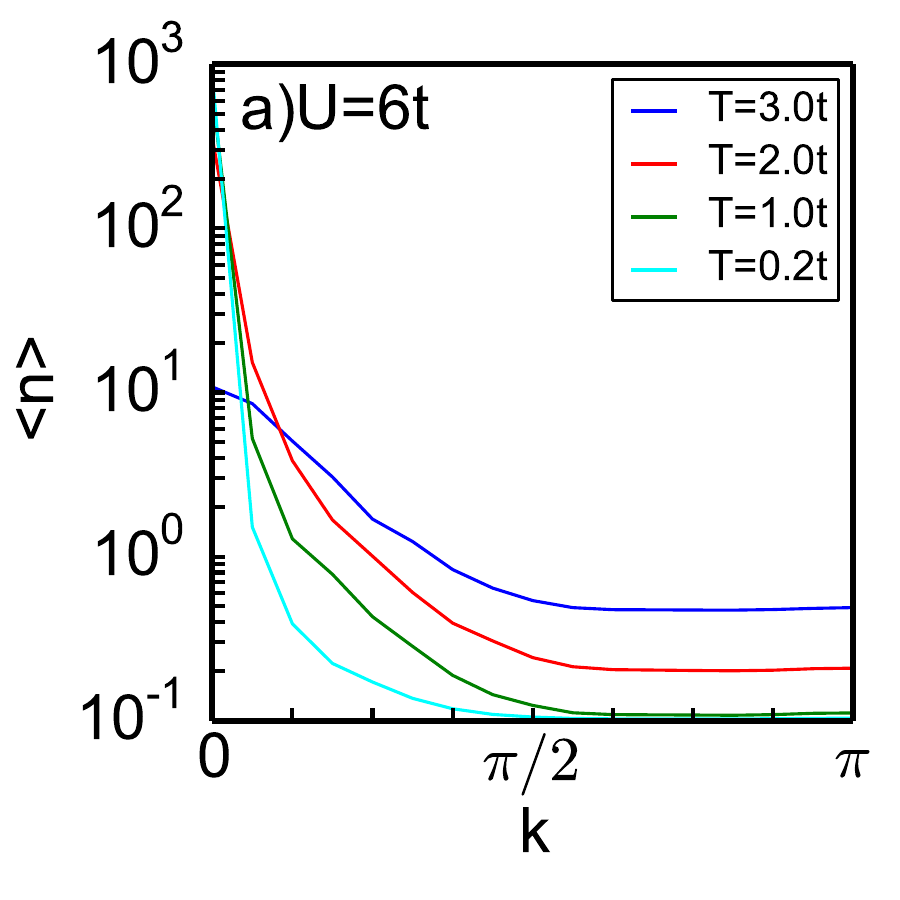}
\includegraphics[angle=0,width=2.7cm,height=4.0cm]{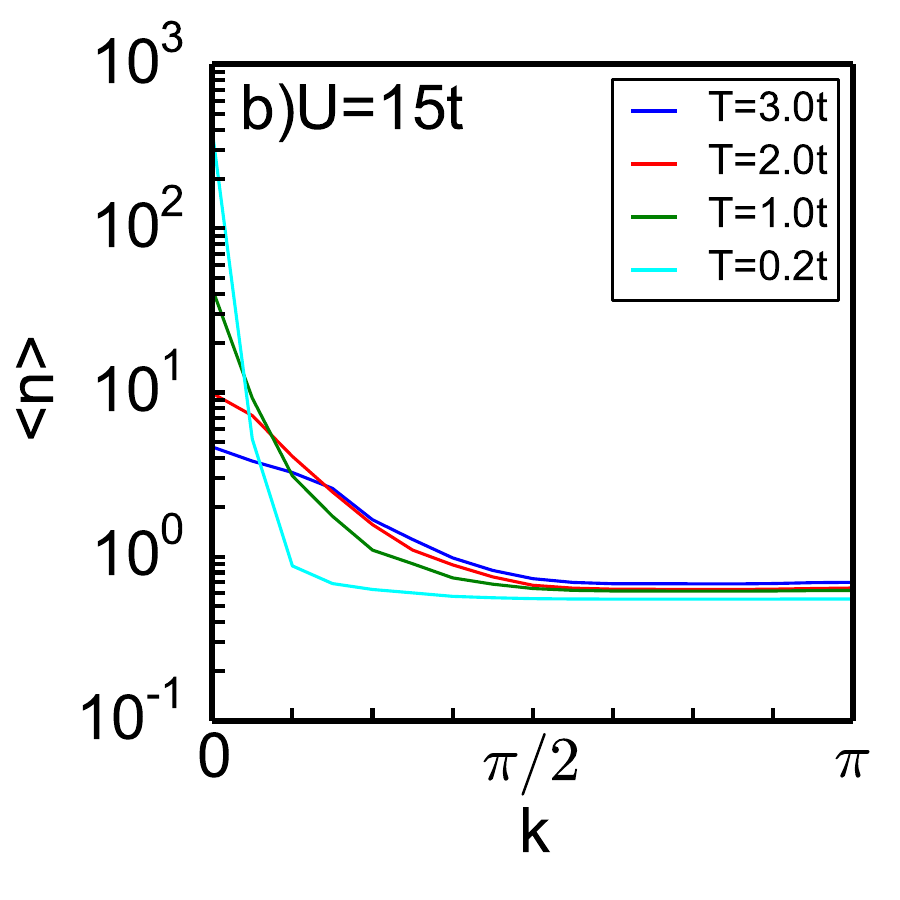} 
\includegraphics[angle=0,width=3.0cm,height=3.8cm]{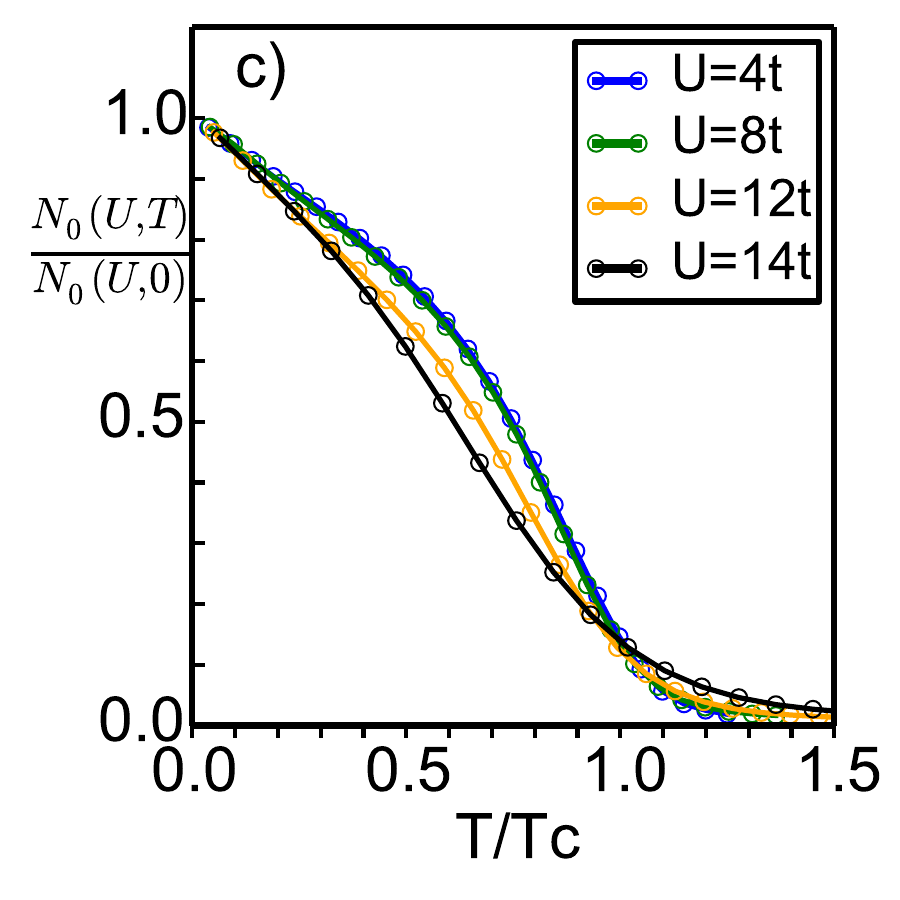}
}
\caption{Color online:
The temperature dependence of the ${\vec k} =0$ occupancy of bosons
for varying $U$. The plot, normalising the occupation  with respect
its $T=0$ value and the temperature with respect to $T_c(U)$
leads to an approximately `universal' result. This, despite the
varying mix of amplitude and phase fluctuations as $U$ varies
from weak to strong coupling.
}
\end{figure}
\begin{figure*}[t!]
\centerline{
\includegraphics[angle=0,width=12.5cm,height=6.5cm]{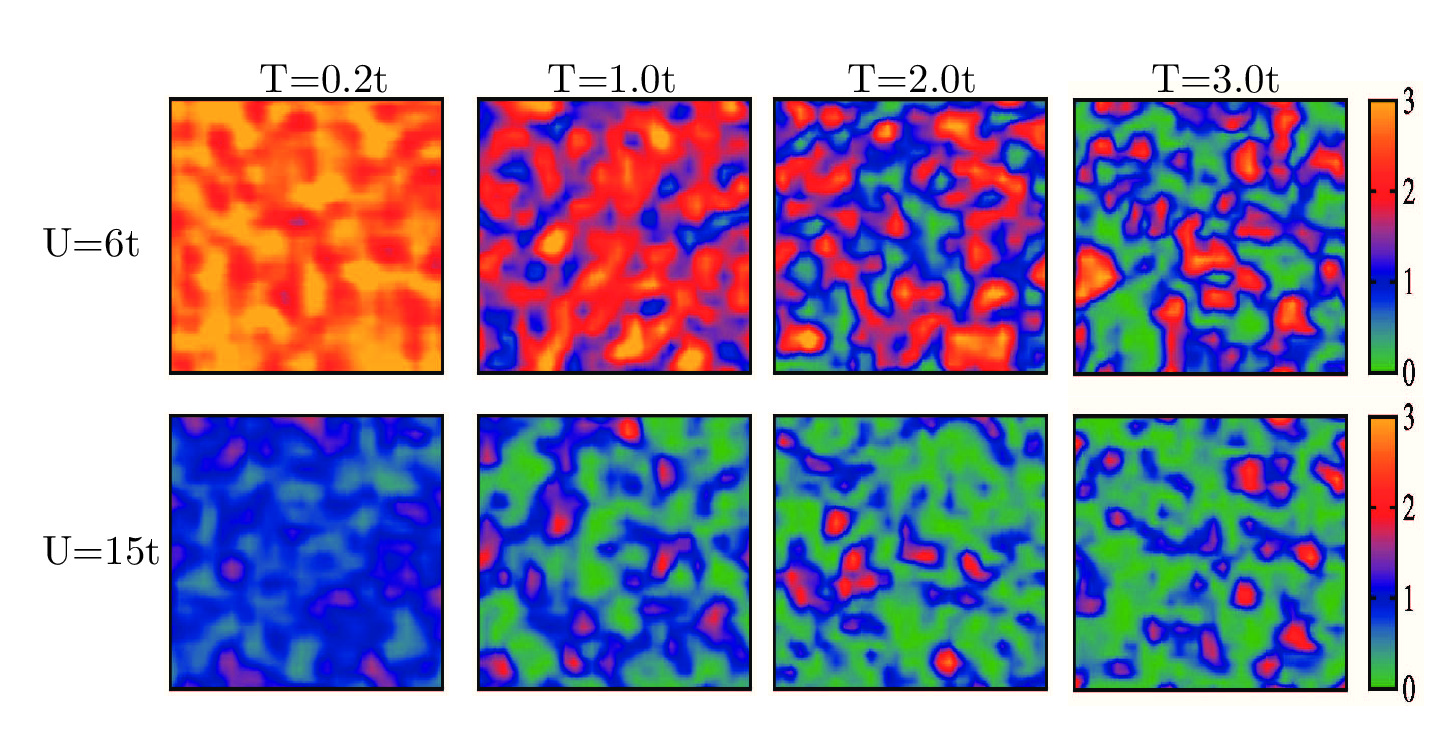}
}
\caption{Color online:
Snapshot of spatial correlations in the superfluid regime.
The maps indicate the correlation $C_i$
of the hybridisation $\Phi_i = \vert \Phi_i \vert e^{ i \theta_i}$
at a site with its four neighbours.  $ C_i= \sum_{\delta}
\vert \Phi_i \vert \vert \Phi_{i + \delta}  \vert
cos(\theta_i - \theta_{i + \delta}) $.
Notice the quasi homogeneous pattern at low $T$ in both
the rows. The connected pattern begins to fragment with
increasing $T$. However, even for $T > T_c$ small
spontaneous clusters with large correlation are present. System size
$32 \times 32$.}
\end{figure*}

\begin{figure}[b]
\centerline{
\includegraphics[angle=0,width=4.4cm,height=4.0cm]{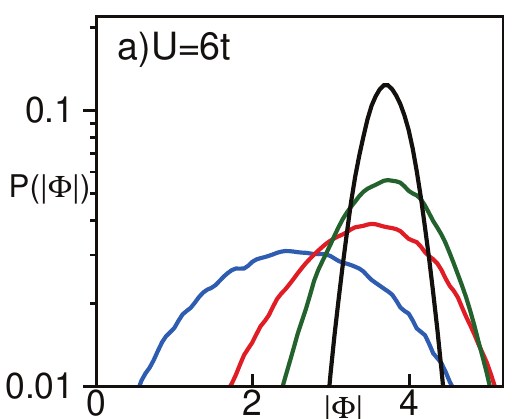}
\includegraphics[angle=0,width=4.cm,height=4.0cm]{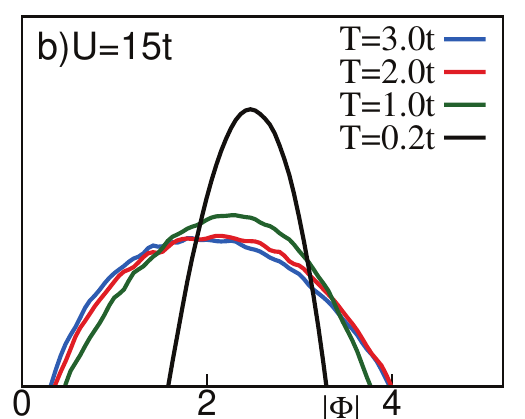}
}
\caption{
Distribution of the magnitude of the local hybridisation.
Panels (a)-(b) have a superfluid ground state.  
The lowest temperature in the data set is $T=0.2t$
at which amplitude fluctuations are already visible. The distributions
appear Gaussian in the weaker $U$ systems.
A later figure shows the mean and variance of these distributions.
}
\end{figure}

\subsubsection{The superfluid phase}

At $T=0$ in the superfluid the hybridisation field is  uniform.
The occupation at $T=0$ consists of a 
peak at ${\vec k} =(0,0)$ and a ${\vec k}$
independent occupancy at ${\vec k} \neq (0,0)$.
The  ${\vec k} =(0,0)$ occupancy, $N_0$, falls with increasing $U/t$
in the ground state, becoming ${\cal O}(1)$ for $U > U_c$.
Fig.6 shows our result for the momentum distribution at $U=6t$ and
$U=15t$ for varying temperature.
We show the quantitative behaviour in Fig.7, but Fig.6 already
reveals that the ${\vec k} = (0,0)$ occupancy falls with $T$
and $U$, and at a given $T$ the `cloud' at $U=15t$ is always
broader than the cloud at $U=6t$.

\begin{figure}[b]
\centerline{
\includegraphics[angle=0,width=4.cm,height=4.0cm]{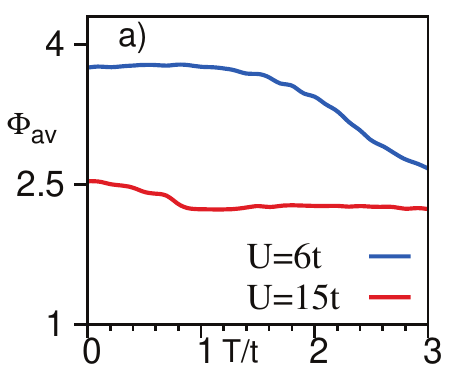}
\includegraphics[angle=0,width=4.cm,height=4.0cm]{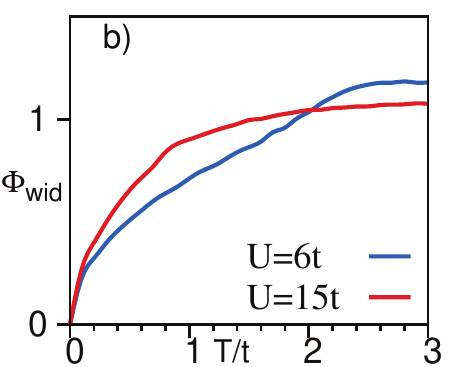}
}
\caption{Color online:
(a)~mean and (b)~standard deviation of the magnitude of the
hybridisation field.
In the superfluid ($U/t \lesssim 16$) the mean falls only weakly
between $T=0$ and  $T_c$, and somewhat faster for $T \gtrsim T_c$.
The magnitude fluctuation
`width' vanishes in all cases as $T \rightarrow 0$,
and grows to a value comparable to the mean by $T \sim 3t$.  }
\end{figure}

Fig.7 quantifies features of the finite $T$ behaviour.
Panels (a) and (b) show the momentum dependence of 
the occupancy along the diagonal $(0,0) \rightarrow
(\pi,\pi)$. The occupancy, plotted on a log scale,
suggests that there is an exponential fall off at
low ${\vec k}$ in the low $T$ regime, tailing off to
a finite ${\vec k}$ independent occupancy at large
momentum.  Crudely, the $T$ dependence seems to follow 
the form 
$$
\langle n({\vec k}) \rangle_T
\sim n_{\infty}(T) + n_0(T) f(k/{\bar k}(T)) 
$$
where $ n_{\infty}(T)$ is the occupancy of the
large momentum states, $n_0(T)$ is the occupancy of
the $(0,0)$ state, 
and ${\bar k}(T) \rightarrow 0$ as $T \rightarrow 0$.
The function $f(x) \rightarrow 0$ as 
$x \rightarrow \infty$ and
$f(x) \rightarrow 1$ as $x \rightarrow 0$.
Panel (c) indicates that the occupancy 
of the ${\vec k} =(0,0)$ mode 
follows an approximate behaviour:
$ { N_0(U,T) \over { N_0(U, 0)}} 
\approx h({T \over T_c(U)}) $
where $h(x)$ can be inferred from Fig.7(c).

Fig.8 shows the correlation $C_i$ for single Monte Carlo snapshots, 
at $U=6t$ (top) and $U=15t$ (bottom).
At the lowest  temperature shown 
there is only weak
amplitude and angular fluctuation, hence
$C_i$ is only weakly inhomogeneous. The second column, $T=t$, 
corresponds to $\sim 0.5T_c$ at $U=6t$ but is 
above $T_c$ at $U=15t$. As a result the top panel still shows   
a connected pattern while the $U=15t$ case shows only small   
correlated droplets in an otherwise uncorrelated background.    
The trend continues to higher $T$, with the correlation length 
for $U=6t$ (which we show later) being larger than that
for $U=15t$ at a given temperature.

\begin{figure*}[t]
{
\includegraphics[angle=0,width=12.0cm,height=6.4cm]{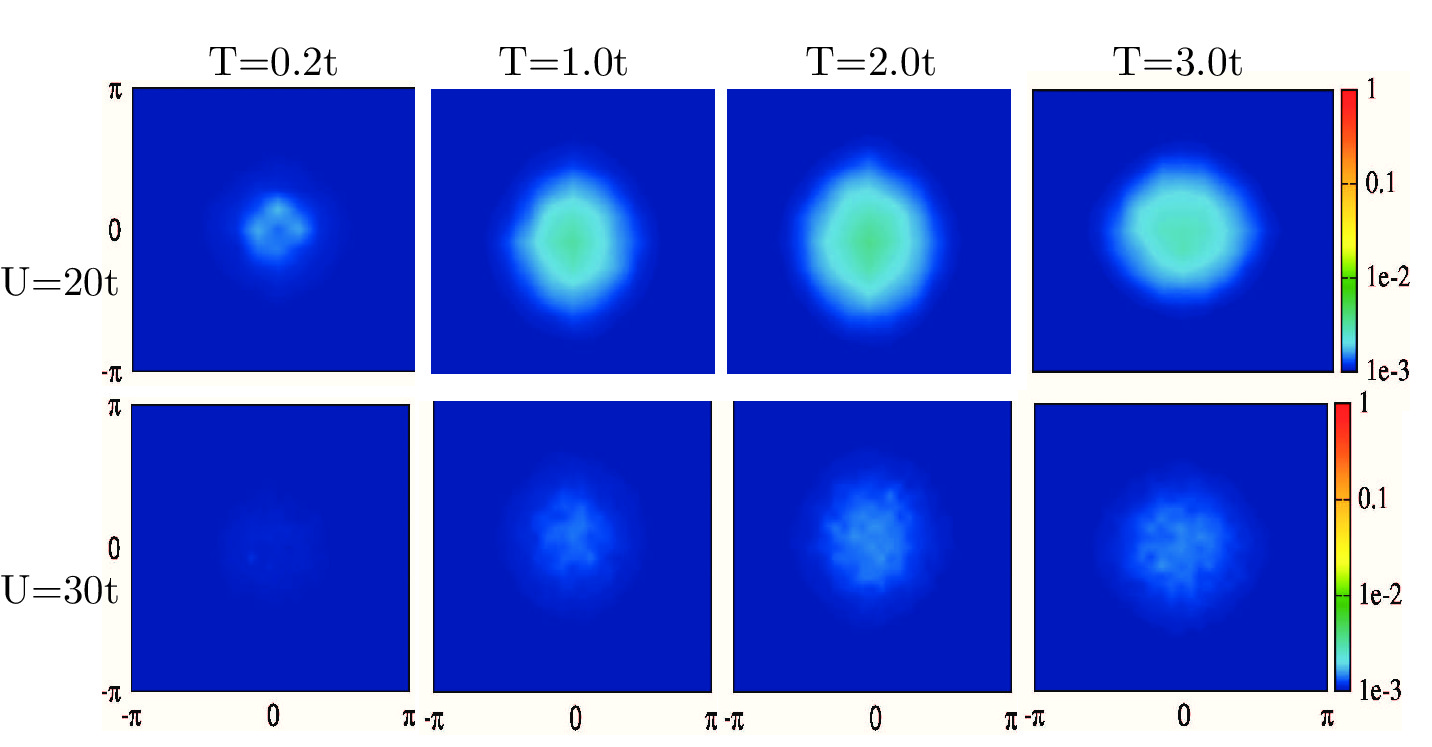}
}
\caption{Color online:
The momentum distribution, $n_{\bf k} =
\langle \langle a^{\dagger}_{\bf k} a_{\bf k} \rangle \rangle$,
for varying interaction strength and temperature in the Mott phase. 
The intensity scale is logarithmic.
We show the thermally excited pattern in the Mott insulator
for $U=20t$ (a weak Mott insulator) and $U=30t$ 
(a strong Mott insulator) in the two rows respectively.  }
\end{figure*}
\begin{figure}[b]
{
\includegraphics[angle=0,width=3.7cm,height=4.0cm]{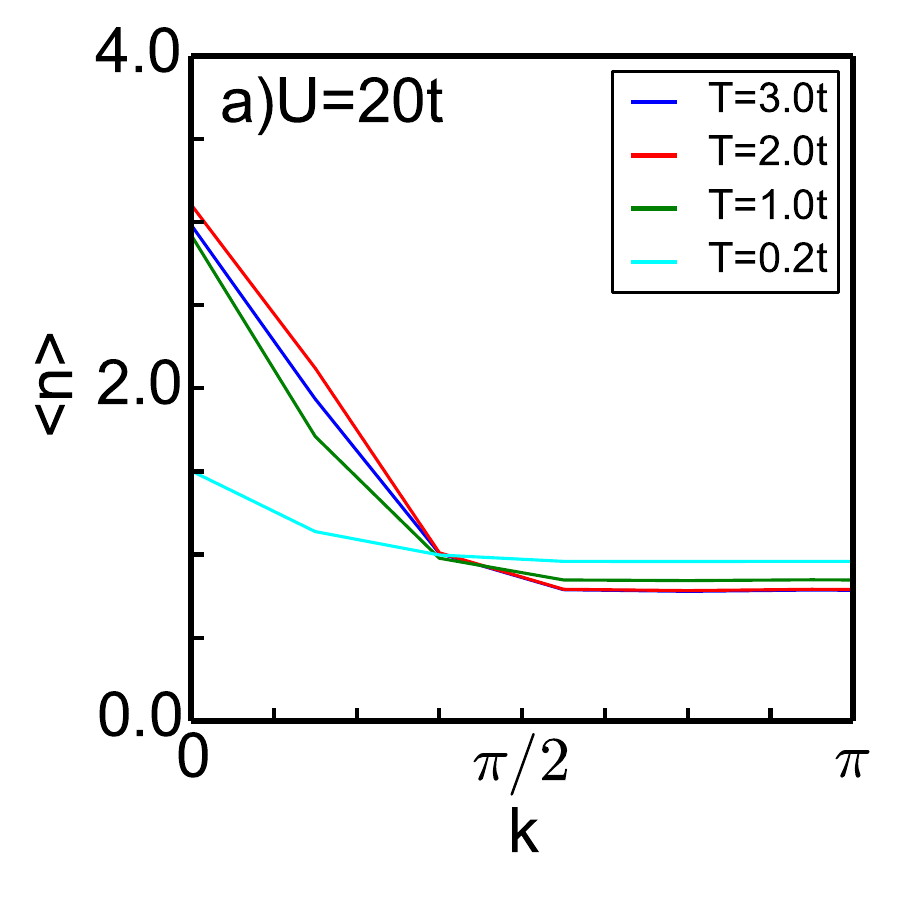}
\includegraphics[angle=0,width=3.7cm,height=4.0cm]{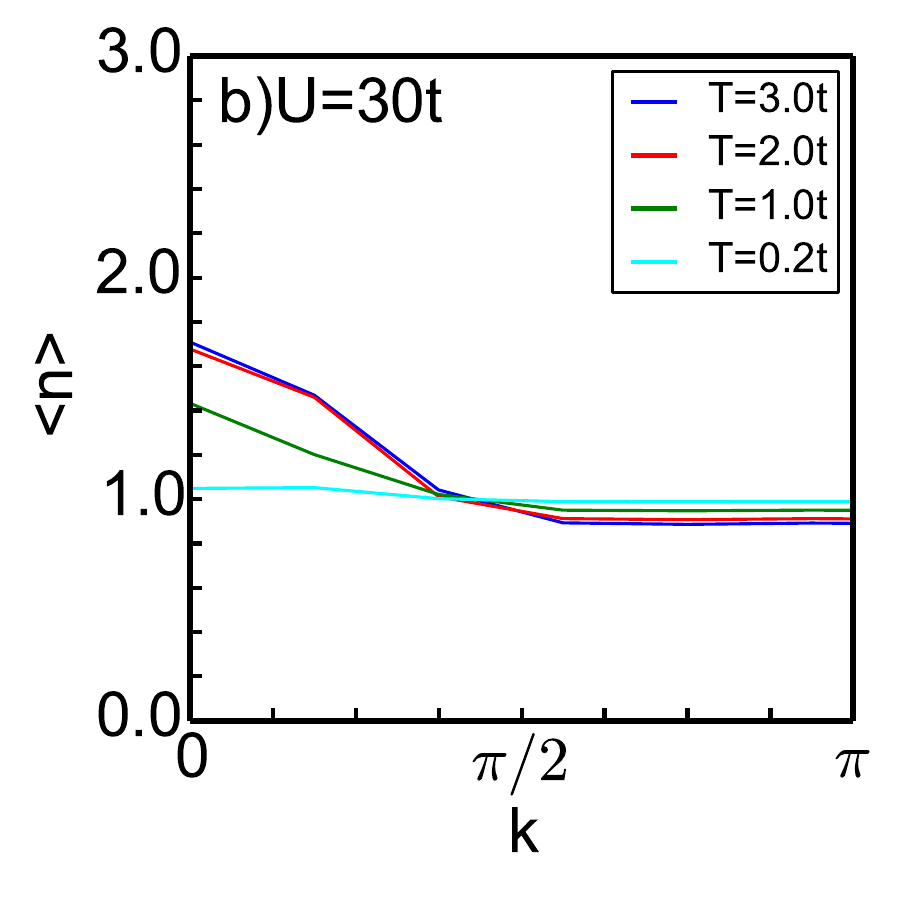}
}
\caption{Color online:
The momentum occupation in the Mott state for varying temperature
for a momentum scan from
${\vec k} \rightarrow (0,0) $ to $(\pi,\pi)$. 
The occupancy is ${\cal O}(1)$ and shows a weak momentum 
dependence in the finite $T$ Mott insulator.
}
\end{figure}
\begin{figure}[b]
\centerline{
\includegraphics[angle=0,width=4.1cm,height=3.5cm]{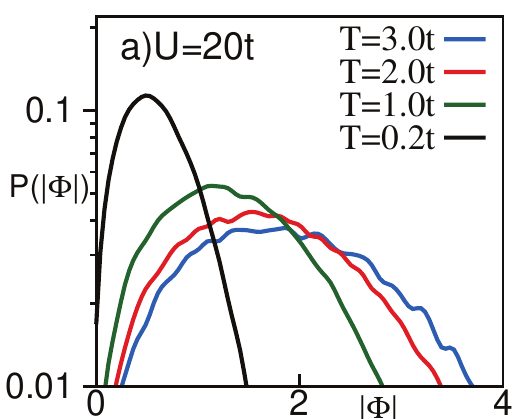}
\includegraphics[angle=0,width=4.1cm,height=3.5cm]{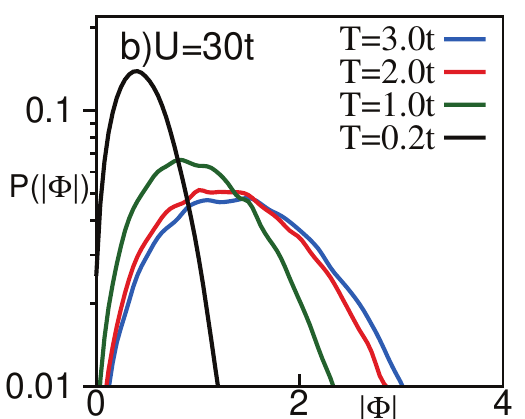}
~
}
\centerline{
\includegraphics[angle=0,width=4.0cm,height=3.5cm]{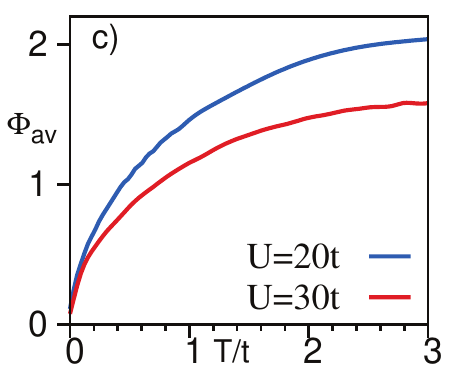}
\includegraphics[angle=0,width=4.0cm,height=3.5cm]{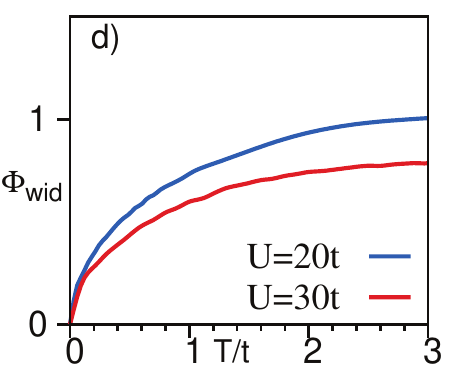}
}
\caption{
(a)-(b) Distribution of $\vert \Phi \vert$ in the Mott insulator.
The lowest temperature is $T=0.2t$
at which amplitude fluctuations are already visible. 
(c)~Mean and (d)~standard deviation of   $\vert \Phi \vert$.
In the Mott phase the mean is zero at $T=0$
but rises quickly attaining a value
$\sim 0.5$ that of the weak $U$ superfluid at $T \sim 3t$. The
width grows as $\sqrt{T}$, to
a value comparable to the mean by $T \sim 3t$.  } 
\end{figure}
\begin{figure*}[t]
\centerline{
\includegraphics[angle=0,width=12.5cm,height=6.4cm]{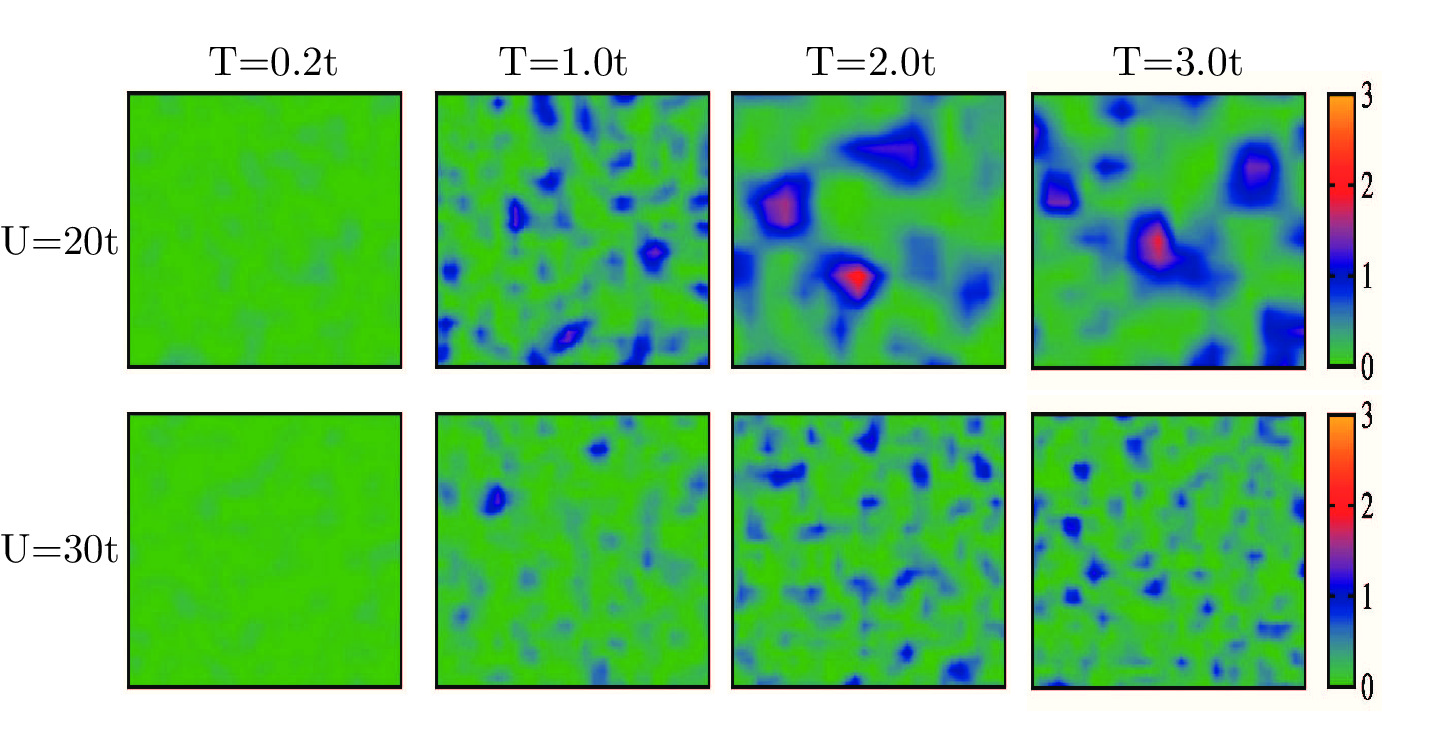}
}
\caption{Color online:
Spatial map indicating the correlation $C_i$
of the hybridisation $\Phi_i = \vert \Phi_i \vert e^{ i \theta_i}$
at a site with its four neighbours.  $ C_i= \sum_{\delta}
\vert \Phi_i \vert \vert \Phi_{i + \delta}  \vert
cos(\theta_i - \theta_{i + \delta}) $.
Notice the vanishingly small $C_i$ in the low $T$
Mott insulator. The Mott insulator also develops small
correlated patches with increasing $T$. System size $32 \times 32$.}
\end{figure*}

We have analysed the auxiliary field backgrounds
in detail.
Fig.9(a) and 9(b) shows the distribution $P(\vert \Phi \vert)$
in the superfluid regime.
At $T=0$ the $\vert \Phi \vert$ is homogeneous across the system and
the distribution is a delta function. 
At finite $T$ we see that $log P((\vert \Phi \vert)$ has a
parabolic character, $ \sim
A(T) - (\vert \Phi \vert - {\bar \Phi}(T))^2/B(T)$,
where $A(T)$ is a normalisation constant, ${\bar \Phi}(T)$
refers the amplitude with 
maximum probability, and $B(T)$ is a measure 
of the width of the distribution.
As is obvious from the plots, for the SF ${\bar \Phi}(T)$ 
falls with increasing $T$, while $B(T)$ rises from zero
as $T$ increases. 
By the time $T \sim T_c$ for both the weak and strongly
interacting superfluids the width of the distribution is
comparable to the mean. Amplitude fluctuations are 
significant all across the high $T$ superfluid.

Fig.10 shows the detailed $T$ dependence of the 
mean and width of the distributions,
computed from the moments.
For the weak coupling SF, with $T_c \sim 2t$, the mean
value starts dropping sharply as $T \rightarrow T_c$, and
the width at $T_c$, in Fig.10(b), is $\sim 0.5$ the mean value.
In the strong coupling case, where $T_c \sim t$, the mean 
value at $T \gtrsim T_c$ is within $20 \%$ of the $T=0$ 
value, although the width is comparable to the weak coupling case.
The width, both at $U=6t$ and $15t$, behaves $\propto \sqrt{T}$ 
at low temperature.

\subsubsection{The Mott insulator}

Fig.11 shows the occupancy 
$\langle n({\vec k}) \rangle$ 
of bosons 
in the weak Mott insulator,  $U=20t$ (top),
and the deep Mott insulator, $U=30t$ (bottom). 
At $T=0$ the hybridisation is zero at all
sites in the Mott insulator and 
$\langle n({\vec k}) \rangle$ 
 is flat over the Brillouin zone.
With increase in $T$, however, a certain structure
becomes visible in the momentum dependence.
A weak peak emerges near ${\vec k} = (0,0)$ and this
feature broadens and picks up intensity at higher
temperature. This sequence is prominent in the
Mott insulator at $U=20t$ than
in the deep Mott state at $30t$. 

This effect arises from the hybridisation
field $\Phi_i$ being generally non zero 
at all sites at finite $T$, 
following a Boltzmann distribution,
and having short range spatial correlation. 
In the next two figures we show the features
of the amplitude distribution - which would be
just $\Phi_i =0$ within mean field theory - and
discuss the spatial correlations later.

Fig.12 shows the momentum occupancy along the diagonal
scan $(0,0) \rightarrow (\pi,\pi)$ in the Mott phase
for varying $T$.
The $T=0$ occupancy within our approximation is flat,
the $\vert \Phi_i \vert$ being zero at all sites. At
finite $T$ the thermally induced hybridisation have a 
spatially correlated pattern, shown later in Fig.14,
which leads to a spatially modulated $G_{ii}$ in the
Monte Carlo configurations. The nanoscale `phase correlated'
patches lead to the weak peak observed in
$\langle n({\vec k}) \rangle$ at small $k$. The
effect expectedly weakens with growing $U/t$.

Fig.13(a)-(b) 
shows the distribution of hybridisation
for $U=20t$ and $U=30t$ while Fg.13 shows the evolution of 
$\Phi_{av}$ and $\Phi_{wid}$ with temperature.
In the Mott phase the hybridisation is 
zero at zero temperature. 
With rise in temperature $P(\Phi)$
picks up weight at non zero $\Phi$.
In panels (c)-(d) both the mean and variance of 
$\Phi$ grow as $T^{\alpha}$ 
with $\alpha \sim 0.5$. The finite temperature Mott state 
has thermally induced particle number fluctuations.
The increase in interaction strength
leads to suppression of number fluctuations.

Fig.14 shows  snapshots showing $C_i$, the
correlation of the 
hybridisation at $R_i$ with its four neighbours in 
the Mott phase. At $T=0$ all $C_i=0$. At finite $T$
there are thermally generated $\Phi_i$, following the
distribution shown in Fig.13, that correlated via 
a coupling that we describe in the next section. 
Since the fluctuation induced 
amplitudes are unlikely to be simultaneously
large over a wide neighbourhood
the patches are small and randomly distributed.

\section{Discussion}

\subsection{Analysing the thermal transition}

Since the thermal behaviour of SPA and PSPA are
similar for $U$ not too close to $U_c$ we focus on
the simpler SPA scheme to suggest a mechanism for the
thermal transition.

The Landau expansion of the SPA functional, assuming a homogeneous
order parameter, $\Phi$, has the form
$$
E(\Phi) = a_0 + a_2|\Phi|^2 + a_4|\Phi|^4 
$$
as we have already seen. The coefficients $a_n$ depend on $\{U,~\mu\}$ and
lead to the minimised value $\Phi_0(U) = 0$ for $U > 24t$ at $n=1$. 
A similar
result, with shifted $U_c$ holds for PSPA. The general finite $T$ form,
where the {$\Phi_i=\sum_jC_{ij}\vert\psi_j\vert e^{i\theta_j}$} has to be
treated like a field, can be obtained via a cumulant expansion of
the free energy shown in the Appendix. This has the general $U(1)$ 
invariant form in terms of $\psi$:
\begin{figure}[b]
\centerline{
\includegraphics[angle=0,width=6cm,height=4.2cm]{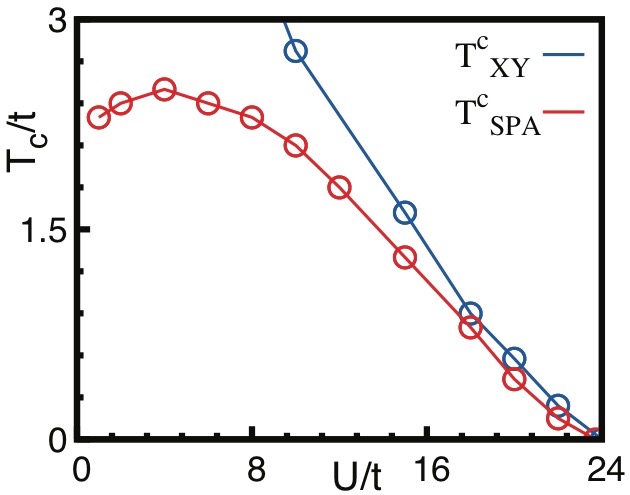}
}
\caption{
Comparison of $T_c$ obtained from the  effective XY model
with $T_c$ from the SPA based Monte Carlo.
For $U \gtrsim 0.5 U_c$, the match is reasonable, at smaller $U$
amplitude fluctuations are significant.
}
\end{figure}

\begin{eqnarray}
F\{ \Phi \} &=& F_0 + F_2 + F_4 + .. \cr
\cr
F_0~~ & = & F\{ \psi=0 \} \cr
\cr
F_2 ~~&=& \sum_{ij} f_{2,ij} \vert \psi_i \vert \vert \psi_j \vert cos(\theta_i - \theta_j)
\cr
F_4 ~~&=& \sum_{ijkl} f_{4,ijkl} 
\vert \psi_i \vert \vert \psi_j \vert \vert \psi_k \vert \vert \psi_l \vert
g(\theta_i,..\theta_l)
\nonumber
\end{eqnarray}
The coefficients $f$ can be calculated in a hopping expansion.

To capture the qualitative physics we simplify as follows:
(i)~We drop the fluctuation of the
amplitudes with $T$, treating them as $\psi_0(U)=\Phi_0(U)/2$, the mean field
value at $T=0$.
This  is consistent with our result
for the finite $T$ mean $\Phi$.
So, we replace all $\vert \psi_i \vert$ by $\Phi_0$.
(ii)~We drop the spatial
dependence of the $g(\theta_i,..\theta_l)$ term.
After this, the only variables
in $F$ are in $F_2$, and the only relevant fluctuations are in the
phase $\theta_i$, as in the XY model.

\begin{figure}[b]
\centerline{
\includegraphics[angle=0,width=4.0cm,height=5.0cm]{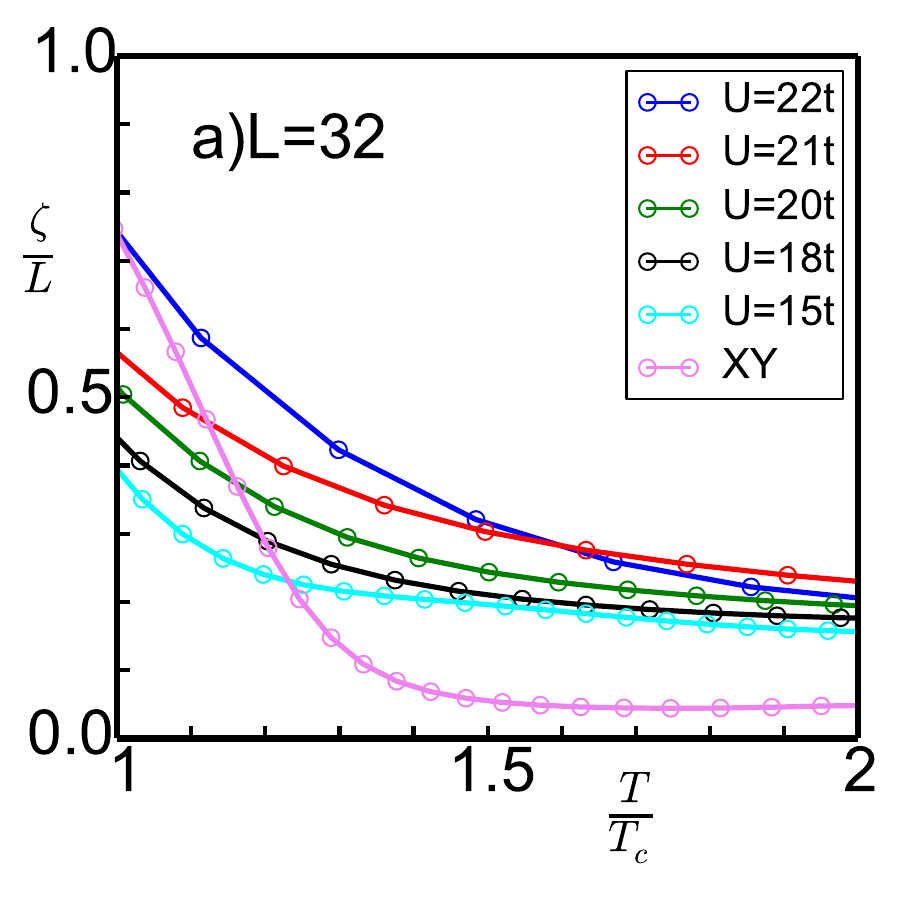}
\includegraphics[angle=0,width=4.0cm,height=5.0cm]{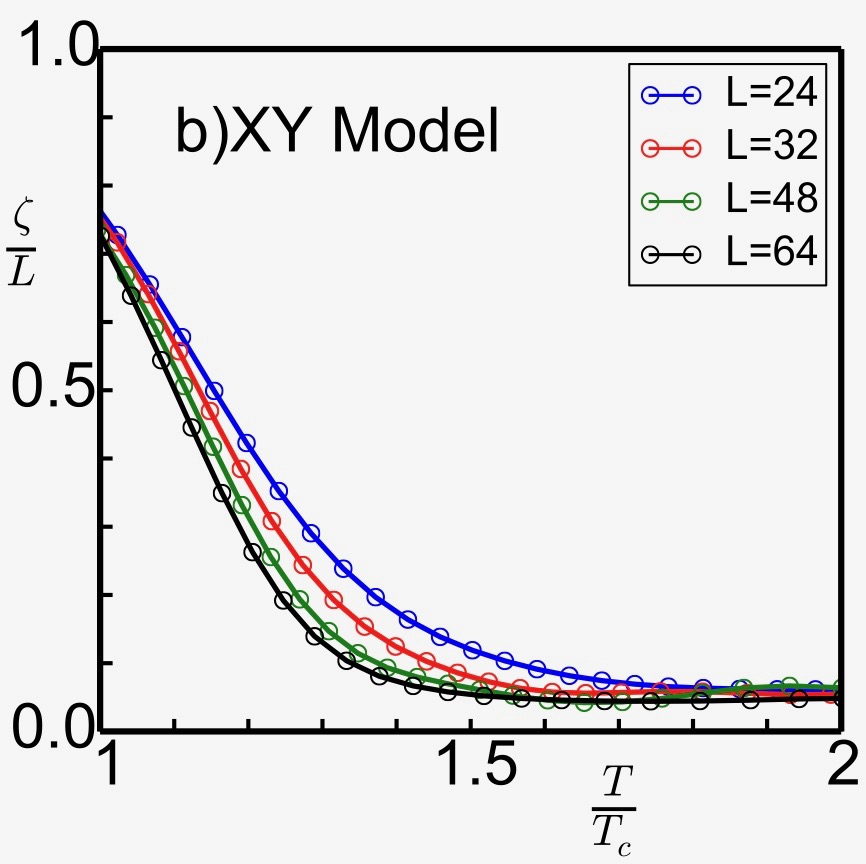}
}
\caption{
(a)~Comparison of correlation scale $\xi(T)/L$ inferred from the
our Bose Hubbard results, plotted with respect to $T/T_c(U)$,
with $\xi(T)/L$ inferred from the nearest neighbour classical 
XY model.
(b)~The system size 
dependence $\xi(T)/L$ for the classical 
XY model, showing that the
qualitative features do not vary much in the $L=24 $ to
$L=32$ window.
}
\end{figure}

\begin{figure}[t]
\centerline{
\includegraphics[angle=0,width=8.3cm,height=5.0cm]{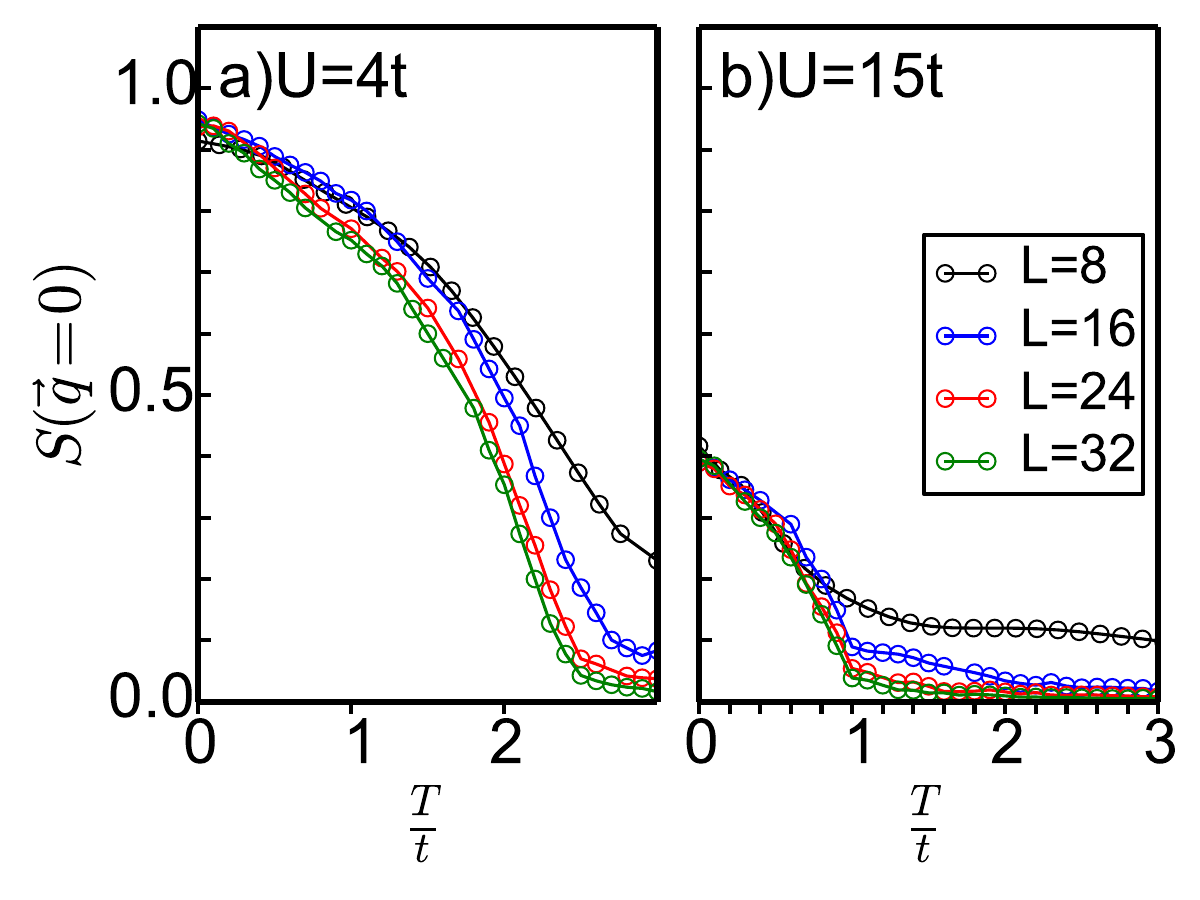}
}
\caption{
Size dependence of the ${\bf q} =(0,0)$ structure factor at
$U=4t,~15t$. At both these $U$ values the function, and the superfluid
onset temperature is almost size independent for $L \gtrsim 24$.
}
\end{figure}

The effective model in this approximation becomes:
$$
F\{ \Phi \} = 
\sum_{ij} f_{2,ij} \Phi^2_0  cos(\theta_i - \theta_j)
$$
The function $f_2$ has a spatial dependence, which we call
$a_{ij}$, and an overall
prefactor $\alpha$ that depends on $U/t,~\mu/t$, {\it i.e}
$ f_{2,ij} = \alpha a_{ij}$. 
This leads to a XY model 
$$
F\{\theta \} = -J\sum_{ij} a_{ij} cos(\theta_i - \theta_j)
$$
where $J = \alpha(U/t,\mu/t) \Phi^2_0(U/t,\mu/t) $. 
The $T_c$ would be controlled by $J$ and the spatial
character of $a_{ij}$ (which depends only on the 
bandstructure).
In Fig.15.
we show a comparison of $T_c$ obtained from the XY model
above with the actual $T_c$ within the SPA scheme. Beyond $U \gtrsim
U_c/2$ the two results match reasonably. 

Fig.16(a) shows the spatial correlation
scale $\xi(T,U)$ extracted
from an analysis of the ${\bf q}$ dependence of the structure
factor about ${\bf q} = (0,0)$.
This analysis is done for the results obtained
using SPA method
since at large $U/t$ ratio thermal transition
scales are well captured by XY model. 

We show 
the comparison of $\xi$ 
for various U values with XY model.
For XY model change in  spatial correlation 
scale is negligible for system size greater than
$L=32$.  We observe that $\xi$ for 
large U values near the $U_c$ matches
with XY result near $T_c$ but high temperature
behaviour do not match with XY model.We also find
correlation scale($\xi$) falls with 
decrease in interaction.

We came across another real space scheme known as
bosonic auxiliary-field Monte Carlo method\cite{Tommaso}.
They have used this method to study hard core 
bosons on square and triangular lattice.

\subsection{Size dependence of our results}

Fig.17 shows the size dependence of our results.  
We did the calculation for four system
sizes $L=16,~24,~32,~40$
for SPA and $L=16,~24,~32$ for
PSPA scheme and show the SPA results for
the ordering feature in the structure factor.
We find that in the superfluid phase,
even at moderate coupling $(U=4t)$, a
system size $L=24$ is large enough to
probe `bulk' behaviour. $L=32$ is hard to
distinguish from $L=24$.
For larger interaction strength,
$U \gtrsim 15t$, $L=16$ seems capture the 
behaviour reasonably. This is related to the
behaviour of the correlation length $\xi(T,U)$ shown in
Fig.16.

\subsection{Results in three dimensions}

We have studied the zero temperature SPA and PSPA
theory in the case of 3D Bose Hubbard model.
At zero temperature
we computed the 
superfluid Mott boundary for unity filling.
The SPA result, as in 2D, matches with
mean field theory, while the PSPA critical point
is almost indistinguishable from the QMC result.
We have not checked the finite temperature results 
in case of 3D
but believe our thermal scales would lie within 20$\%$
of QMC results.

\subsection{Extensions of the method}

Our technique is general enough to handle any kind 
of diagonal disorder as in case of diagonal disorder the
only change is in the exactly handled local
Hamiltonian. 
We can also handle hopping disorder 
through a change in coupling between
the auxiliary field and the bosons.
We can of course handle smooth potentials like traps.
We can also study spin-orbit coupling or artificial 
gauge fields in 
 multispecies bosons as we will separately report.
Finally, given the equilibrium classical 
backgrounds we can compute Green's functions of the
Bose theory via a strong coupling expansion. 
We will present results on this soon.

\section{Conclusions}

We have computed the phase diagram of the two dimensional Bose Hubbard model
at integer filling. The results, based on a classical approximation to the
`hybridisation field', and a quantum correction on it, yield transition
scales, $T_c(U)$, 
that compare favourably with full quantum Monte Carlo results.
The simpler version of the method, called SPA, has a computational cost
that scales linearly with system size, $N$, with a coefficient
$\sim N_c N_b^3$, where $N_c$ is a cluster size $\sim 100$ and $N_b$ 
is the number of atomic states retained per site.
The method is framed in real space, unlike DMFT. 
As a result it can capture the spatial amplitude and phase fluctuations
on a specific lattice, or disordered background, or in the presence of a trap.
It is also framed in   `real time', unlike QMC, and, as we will separately
present, allows access to spectral information about the system.

We acknowledge use of the High Performance Computing Facility at
HRI. Abhishek Joshi thanks the Infosys Foundation for support.
AJ acknowledges fruitful discussions with Sauri Bhattacharyya,
Arijit Dutta and Samrat Kadge.

\appendix

\section{Derivation of PSPA}

 $Z=\int \prod_i D[\psi,{\psi^*}] ~D[b,\overline b] \text{e}^{-(S+S_b)}$\\\\
 $Z_0$ is the adiabatic partition function\\\\
 $Z_0=\int D[b,\overline b] \text{e}^{-S}= Tr[exp(-\beta H')]$\\\\
 $e^{-S_b}$ is expanded to rewrite Z as\\\\
 $Z=\int  D[\psi, 
{\psi^*}]D[b,\overline b] \text{e}^{-S}
(1-S_b+\frac{S_b^2}{2!}+..)$\\\\
$Z=\int  D[\psi,
 {\psi^*}]~Z_0(1-\frac{\int D[b,\overline b] 
{e}^{-S}S_b}{Z_0} 
+\frac{\int D[b,\overline b]
 {e}^{-S}S_b^2}{2!~Z_0}+.....)\\\\
 S_b=-\sum\limits_{i,j,n\neq0}  t_{i,j}\overline b_n
b_{{j},n}-\sum\limits_{i,j}B_{i,j}\overline b_{i,0}b_{{j},0}\\$

$=-\sum\limits_{i,j,n} t_{i,j}\overline b_{i,n}b_{{j},n}
+\sum\limits_{i,j}(t_{i,j}-B_{ij})\overline b_{i,0}b_{{j},0}\\\\
\beta X=\frac{\int D[b,\overline b] \text{e}^{-S}S_b}{Z_0}\\$

X includes effects of zero frequency part of positive band\\

$X=\frac{\sum\limits_{i\neq j}
(t_{ij}-B_{i,j})~\text{Tr}[e^{-\beta H'_i}~  b_i^{\dagger}]
\text{Tr}[e^{-\beta H'_j}~  b_j]}{\text{Tr}
[e^{-\beta H'_i}]\text{Tr}[e^{-\beta H'_j}]}$\\

$-\sum\limits_i B_{ii}(  \sum
\limits_n e^{-\beta E^i_n }|<n| b_i^{\dagger}|n>|^2$\\

$-\frac{1}{\beta}\sum\limits_{n\neq m}
<n|b_i^{\dagger}|m><m| b_i|n>\frac{e^{-\beta E^i_n}
-e^{-\beta E^i_m}}{(E^i_m-E^i_n)})$

$\beta  Y=\frac{\int D[b,\overline b]
 \text{e}^{-S}S_b^2}{2!~Z_0}-\frac{\beta^2X^2}{2!}- (small~ corr)$\\

Y include second order correction due non zero frequency
part of  to kinetic term.It has anomalous and normal 
contribution\\

$Y=\sum\limits_{i,j}t^2/2(F1_{i,j;i,j+1}
+F1_{i,j;i+1,j}+F2_{i,j;i+1,j}\\~~~~~~~~~+F2_{i,j;i,j+1})$\\

$|u>,|v>$ are eigenvector of $H'$ at site index i ,j
$|p>,|q>$ are eigenvector of $H'$ at site index k,m
F1 is anomalous contribution at  second order \\

$F1_{i,j;k,m}=\sum\limits_{\substack
{u\neq v\\\ p\neq q}}(\chi_{uvpq}+\overline\chi_{uvpq})(A_{i,j;k,m}^{uvpq}+\frac{1}{4\beta}
B_{i,j;k,m}'^{uvpq})$

$\chi_{uvpq}=<u|b_{i,j}^\dagger|v>
<v| b_{i,j}^\dagger|u> <p|b_{k,m}|q>\\\hspace*{1.5cm}<q| b_{k,m}|p>\\$

$\overline\chi_{uvpq}=<u| b_{i,j}|v><v| b_{i,j}|u>
 <p| b_{k,m}^\dagger|q>\\\hspace*{1.5cm}<q|b_{k,m}^\dagger|p>\\$

F2 is normal contribution at second order\\

$F2_{i,j;k,m}=\sum\limits_
{\substack{u\neq v\\\ p\neq q}}(\chi'_{uvpq}+\overline\chi'_{uvpq})(A_{i,j;k,m}^{uvpq}
+\frac{1}{4\beta}B_{i,j;k,m}'^{uvpq})\\$

$\chi'_{uvpq}=<u|b_{i,j}^\dagger|v>
<v|b_{i,j}|u> <p|b_{k,m}|q>\\\hspace*{1.5cm}<q|b_{k,m}^\dagger|p>\\$

$\overline\chi'_{uvpq}=<u| b_{i,j}|v><v|b_{i,j}^\dagger|u> 
<p|b_{k,m}^\dagger|q>\\\hspace*{1.5cm}<q| b_{k,m}|p>\\$

$A_{i,j;k,m}^{uvpq}=\frac{e^{-\beta(\epsilon_q^{k,m} 
+\epsilon_v^{i,j})}-e^{-\beta (\epsilon_u^{i,j}+\epsilon_p^{k,m})}}
{\epsilon_u^{i,j}+\epsilon_p^{k,m}-\epsilon_q^{k,m}-
\epsilon_v^{i,j}} if~ den\neq 0 $\\

$else~~A_{i,j;k,m}^{uvpq}=
\beta e^{-\beta(\epsilon_q^{k,m} +\epsilon_v^{i,j})}\\$

$B_{i,j;k,m}'^{uvpq}=
\frac{e^{-\beta \epsilon_v^{i,j}}-e^{-\beta \epsilon_u^{i,j}}}
{\epsilon_u^{i,j}-\epsilon_v^{i,j}}~~~
\frac{e^{-\beta \epsilon_q^{k,m}}-e^{-\beta \epsilon_p^{k,m}}}
{\epsilon_q^{k,m}-\epsilon_p^{k,m}}\\$

$Z=\int  D[\psi,
{\psi^*}]e^{-\beta F}(1-\beta X+..+\frac{\beta^2 X^2}{2!}
+\beta Y+..\\$
\hspace*{1cm}+higher order (diagrams+powers of X and Y ))\\

$Z\approx\int D[\psi,
{\psi^*}]e^{-\beta( F+X-Y)}\\$

$F=\frac{-1}{\beta}log(Tr[exp(-\beta H')])\\$
This is cumulant expansion. One exponentiate the 
series and we keep terms only  till second order
in the corrected free energy.\\

\section{Approximate SPA functional at zero temperature}

The full partition function is given by\\

$Z_{SPA}= \int 
D[ \psi, {\psi^*}] ~D[b,\overline b]~ e^{-S}$\\

$S=S_0+S_{pert}$\\

$S_0= \int_0^{\beta} d\tau [\sum_i
{\overline b_i}~ (\partial_{\tau} - \mu) b_i+\psi_i^*\psi_i+{U \over 2}  \overline b_i 
b_i(\overline b_i b_i-1)] \\$

$S_{pert}= \int_0^{\beta} d\tau\sum_{ij}-C_{ij}(\overline b_i(\tau)\psi_j+h.c)
\\$

At zero temperature the minimum energy solution is where
$\psi$ is uniform. So we take $\psi_j$=$\psi$. Now\\

$S_{pert}= \int_0^{\beta} d\tau\sum_{i}(-2\sqrt{t}~\overline b_i(\tau)\psi_j+h.c)\\$

since $\sum_j C_{ij}$=2$\sqrt{t}$\\

$Z_{SPA}=  \int 
D[\psi,{\psi^*}] 
~D[b,\overline b]~ e^{-S_0}(1-S_{pert}+\frac{S_{pert}^2}{2!}+..) \\$

odd power in $S_{pert}$ is zero due to number conservation
The cumulant expansion till fourth order 
and dropping higher order terms and exponentiating one gets\\

$Z_{SPA}(T=0)\approx \int ~ (d\psi d\psi^*)^N~ e^{-\beta E(\psi)}\\$

$\frac{E(\psi)}{N}=a_0+a_2|\psi|^2+a_4|\psi|^4\\$

The ground state phase boundary is obtained by minimizing
E$(\psi)$ , where a$_0$, a$_2$ and a$_4$  is defined in terms of atomic 
green function.N is the number of sites.\\

$a_0=[U/2n(n-1)-\mu n]$\\

$a_2=(1+4~t~G_{ii}(0))\\$

$a_4={-4~t^2}G_{ii}^{2c}\\$

G$_{ii}$(0) and G$_{ii}^{2c}$ is defined as below \\

$G_{ii}(0)=\int_0^{\beta}d\tau-<T_\tau b(\tau)b^\dagger(0)>\\$

$G_{ii}^{2c}=\int_0^{\beta} d\tau_1 d\tau_2 d\tau_3
<T_{\tau}b(\tau_1)b(\tau_2)b^\dagger(0)b^{\dagger}(\tau_3)>\\$
$\hspace*{1cm}-2\beta G_{ii}(0)^2\\$

At zero the temperature SPA functional is same as the mean field
functional

\section{Approximate PSPA functional at zero temperature}
The full partition function is given by\\

$Z=\int \prod_i D[\psi,
 {\psi^*}] ~D[b,\overline b] \text{e}^{-(S_b+S_{pert}+S_b)}\\$

At zero temperature the minimum energy solution 
is where $\psi$ is uniform.\\

$Z(T=0)=\int~  (d\psi d\psi^*)^N D[b,\overline b]e^{-S_0}(1-(S_b+S_{pert})\\$
$\hspace*{2cm}+\frac{(S_b+S_{pert})^2}{2!}+...)\\$

After integrating out bosons order by order one gets\\

$=\int  (d\psi d\psi^*)^N<e^{-S_0}>(1+\frac{<S_{pert}^2>}{2!}+\frac{<S_{pert}^4>}{4!}\\$

$+\frac{<S_{b}^2>}{2!}+\frac{<S_{pert}^2S_b^2>}{4}+\text{higher order terms})\\$

One can do cumulant expansion and drop higher terms.
Above series can be approximated by\\

$Z_{PSPA}(T=0)\approx\int~(d\psi d\psi^*)^N e^{-\beta E_{corr}(\psi)}\\$

$E_{corr}(\psi)=E(\psi)+\delta E(\psi)\\$

$E(\psi)=-\frac{1}{\beta}(log(e^{-<S_0>})+\frac{<S_{pert}^2>}{2!}+\frac{<S_{pert}^4>}{4!})\\$

$\delta E(\psi)=-\frac{1}{\beta}(\frac{<S_{b}^2>}{2!}+\frac{<S_{pert}^2S_b^2>}{4})\\$

This is the approximated PSPA functional whose terms are explained as below\\

$\frac{E_{corr}(\psi)}{N}=(a_0+\delta a_0)+(a_2+\delta a_2)|\psi|^2+a_4|\psi|^4\\$

$\frac{E(\psi)}{N}=(a_0)+(a_2)|\psi|^2+a_4|\psi|^4\\$

where a$_0$, a$_2$, a$_4$ are described as  in previous section.
$\frac{\delta E(\psi)}{N}=\delta a_0+\delta a_2 |\psi|^2\\$

$\delta a_0 $ and $\delta a_2$ are obtained from atomic green function\\

$\delta a_0=-\frac{2}{\beta}~\int d\tau_1d\tau_2~ t^2 G_{ii}(\tau_1,\tau_2)G_{jj}(\tau_2,\tau_1)\\$

$\delta a_2=-\frac{8}{\beta} (~\int_0^\beta d\tau_1 
d\tau_2 d\tau_3 d\tau_4 G^2(\tau_1,\tau_2;\tau_3,\tau_4)
G(\tau_2,\tau_3)\\$
$\hspace*{1cm}+G^2(i\omega_1=0,i\omega_2=0;i\omega_3=0,i\omega_4=0)
G(i\omega_5=0,i\omega_6=0))\frac{1}{\beta^2}\\$

where G$^2$ is two particle green function given as below
and G is single particle green function\\

$G^2(\tau_1,\tau_2;\tau_3,\tau_4)=
<T_{\tau}b(\tau_1)b(\tau_2)b^\dagger(\tau_3)b^{\dagger}(\tau_4)>\\$

$G(\tau_1,\tau_2)=-<T_\tau b(\tau_1)b^\dagger (\tau_2)>\\$

The bosonic SPA is a `single site' theory, albeit with a spatially
correlated hybridisation field. The PSPA 
incorporates  effects due to tunneling of particles to other
sites.	To lowest order PSPA corrects the $|\psi|^2$ term
and leads to shifting of the phase boundary.


\begin{thebibliography}{99}
\bibitem{Bloch} M. Greiner, O. Mandel, T. Esslinger, 
T. W. H\"{a}nsch, and I. Bloch,
Nature (London) {\bf 415}, 39 (2002).

\bibitem{Fort} D. Cl\`{e}ment, N. Fabbri, L. Fallani,
C. Fort, and M. Inguscio,
Phys. Rev. Lett. {\bf 102}, 155301 (2009).

\bibitem{Ernst} P. T. Ernst, S. G\"{o}tze,
J. S. Krauser, K. Pyka, D. Lühmann,
 D. Pfannkuche and K. Sengstock,
Nature Physics {\bf 6}, 56 (2010).

\bibitem{Klaus} U. Bissbort, S. G\"{o}tze,
 Y. Li, J. Heinze, J. S. Krauser,
Malte Weinberg, C. Becker, K. Sengstock,
and W. Hofstetter,
Phys. Rev. Lett. {\bf 106}, 205303 (2011).

\bibitem{Immanuel Bloch} I. Bloch,
Nature Physics {\bf 1}, 23 (2005).

\bibitem{Zwerger} I. Bloch, J. Dalibard,
 and W. Zwerger,
Rev. Mod. Phys. {\bf 80}, 885 (2008).


\bibitem{Sansone} B. Capogrosso-Sansone,
S. G. S\"{o}yler, N. Prokof$'$ev,
and B. Svistunov,
Phys. Rev. A, {\bf 77}, 015602 (2008).

\bibitem{Capogrosso} B. Capogrosso-Sansone,
 N. V. Prokof$'$ev, B. V. Svistunov,
Phys. Rev. B {\bf 75}, 134302 (2007).

\bibitem{Sheshadri} K. Sheshadri, H. R. Krishnamurthy, 
R. Pandit, T. V. Ramakrishnan,
Europhys. Letters, {\bf 22}, (4), 257 (1993).

\bibitem{Fisher} M. P. A. Fisher, P. B. Weichman, 
G. Grinstein, and D. S. Fisher,
Phys. Rev. B {\bf 40}, 546 (1989).

\bibitem{Monien} N. Elstner, H. Monien,
Phys Rev B {\bf 59},12184 (1999).

\bibitem{Freericks} J. K. Freericks, H. Monien,
Europhys. Letters, {\bf 26}, (7), 545 (1994).

\bibitem{dos}F. E. A. dos Santos, A. Pelster,
Phys. Rev. A {\bf 79}, 013614 (2009).

\bibitem{Anirban} A. Dutta, C. Trefzger,
 and K. Sengupta,
Phys. Rev. B {\bf 86}, 085140 (2012).

\bibitem{Dirk}
Dirk-Sören Lühmann
Phys. Rev. A 87, 043619 


\bibitem{Knap} M. Knap, E. Arrigoni, and
 W. von der Linden,
Phys. Rev. B {\bf 81}, 235122 (2010).

\bibitem{Michael} M. Knap, E. Arrigoni, 
and W. von der Linden,
Phys. Rev. B {\bf 83}, 134507 (2011).


\bibitem{Stoof}D. B. M. Dickerscheid, D. van Oosten,
 P. J. H. Denteneer, and H. T. C. Stoof,
Phys. Rev. A {\bf 68}, 043623 (2003).

\bibitem{Lu} X. Lu, J. Li and Y. Yu,
Phys. Rev. A {\bf 73}, 043607 (2006).

\bibitem{SBO}A. S. Sajna, T. P. Polak, R. Micnas, and
 P. Ro\.{z}ek, Phy. Rev. A {\bf 92}, 013602 (2015).


\bibitem{Volhardt} K. Byczuk and D. Vollhardt,
Phys. Rev. B {\bf 77}, 235106 (2008).

\bibitem{Tong} Wen-Jun Hu and Ning-Hua Tong,
Phys. Rev. B {\bf 80}, 245110 (2009).

\bibitem{Anders} P. Anders, E. Gull, L. Pollet,
 M. Troyer and Philipp Werner,
Phys. Rev. Lett. {\bf 105}, 096402 (2010).

\bibitem{Pollet} D. H\"{u}gel and L. Pollet,
Phys. Rev. B {\bf 91}, 224510 (2015).

\bibitem{Kampf} A. P. Kampf and G. T. Zimanyi,
 Phys. Rev. B {\bf 47}, 279 (1993).


\bibitem{HTC} D. van Oosten, P. van der Straten, and H. T. C. Stoof,
 Phys. Rev. A {\bf 63}, 053601 (2001). 


\bibitem{Sengupta} K. Sengupta and N. Dupuis,
Phys. Rev. A {\bf 71}, 033629 (2005).

\bibitem{Negele}B. Lauritzen and J. W. Negele
Phys. Rev. C {\bf 44}, 729 

\bibitem{Bertsch}
B. Lauritzen and G. Bertsch
Phys. Rev. C {\bf 39}, 2412 

\bibitem{Attias}
H. Attias, Y. Alhassid 
Nuclear Physics A 625 (1997) 565-597 

\bibitem{Tommaso}
Daniele Malpetti and Tommaso Roscilde
Phys. Rev. Lett. 119, 040602 


\end{thebibliography}
\end{document}